\documentclass[11pt,a4paper]{article}
\usepackage[novbox]{pdfsync}
\usepackage{enumerate}
\usepackage{enumitem}
\usepackage{tabularx}
   \usepackage[margin=0.5in]{geometry}
\usepackage[utf8]{inputenc}
\usepackage{graphicx}
\graphicspath{{Figures/}}
\usepackage[T1]{fontenc}
\usepackage{fancyhdr, float}
\usepackage{mathrsfs,amssymb,amsmath,amsfonts,booktabs,array,xcolor,url,cite,
color,multirow,multicol}
\usepackage{slashbox}
\usepackage{stfloats}
\usepackage{extarrows}
\usepackage{textcomp}
\usepackage{ulem}
\usepackage{cases}
\usepackage{bm}
\usepackage{arydshln}
\usepackage{hyperref}\usepackage{array,subcaption}\usepackage{verbatim}
\usepackage[font=footnotesize]{caption}
\usepackage[all]{xy}
\usepackage{listings}
\usepackage{tikz, pgf,graphicx, pgfplots, color, xcolor}
	\pgfplotsset{compat=1.12} 
\usepackage{tikz-3dplot}
\usepackage{tikz-cd}
	\usetikzlibrary{positioning}

\newtheorem{theorem}{Theorem}
\newtheorem{remark}{Remark}
\newtheorem{proposition}{Proposition}
\newtheorem{definition}{Definition}
\newtheorem{corollary}{Corollary}
\newtheorem{example}{Example}
\newtheorem{lemma}{Lemma}
\newtheorem{assumption}{Assumption}

 \def\eeD{\end{definition}} \def\beD{\begin{definition}}
\def\beR{\begin{remark}} \def\eeR{\end{remark}}
\def\beL{\begin{lemma}} \def\eeL{\end{lemma}}
\def\beC{\begin{corollary}
  }\def\eeC{\end{corollary}}
  \def\beT{\begin{theorem}}\def\eeT{\end{theorem}}
  \def\beP{\begin{proposition}} \def\eeP{\end{proposition}}
\def\beXa{\begin{example}} \def\eeXa{\end{example}}
\def\beA{\begin{assumption}} \def\eeA{\end{assumption}}
\def\im{\item}\def\com{compartment}     \def\vi{\overset{\rightarrow}{\i}}   \def\F{F}

      \def\mV{{\mathcal V}}
  \def\Prf{{\bf  Proof:}} \def\PF{Perron Frobenius } \def\NGM{next generation matrix}
\def\brn{basic reproduction number}\def\DFE{disease-free equilibrium}
     \def\sd{\s_{dfe}}

 \newcommand{\e}{\;\mathsf e}

\renewcommand{\i}{\;\mathsf i}
\renewcommand{\r}{\; \mathsf r}

\def\fe{for example} 
\def\BEN{\begin{enumerate}}  \def\BI{\begin{itemize}}
\def\EEN{\end{enumerate}}   \def\EI{\end{itemize}}
\newcommand{\beq}{\begin{eqnarray}
    }
\def\eeq{\end{eqnarray}}
   \newcommand{\be}[1]{\begin{equation}\label{#1}}
\newcommand{\ee}{\end{equation}}
\def\bea{\begin{eqnarray*}} \def\im{\item} \def\Lra{\Longrightarrow}  \def\eqr{\eqref}  
\def\no{\nonumber} 

\newcommand{\R}{\mathbb{R}}
\newcommand{\N}{\mathbb{N}}

 \def\sd{\s_{dfe}}\def\al{\al}\def\sec{\section}
\def\b{\beta} \def\g{\gamma}    \def\de{\delta}
\def\z{\zeta}   
\def\ep{\epsilon}  \def\l{\lambda}
   \def\al{\al}
\def\L{b} \def\mR{\mathcal R} \def\mI{\mathcal I}\def\fr{\frac}
\def\bc{\begin{cases}
  }
\def\ec{\end{cases}}
\def\bea{\begin{eqnarray*}}\def\ssec{\subsection} \def\sssec{\subsubsection}

\def\eea{\end{eqnarray*}} 
   \def\I{\infty}
\def\no{\nonumber}\def\ngm{next generation matrix}\def\mF{\mathcal F}  
\long\def\symbolfootnote[#1]#2{
\begingroup
\def\thefootnote{\fnsymbol{footnote}}\footnote[#1]{#2}
\endgroup}
\def\fn{\symbolfootnote}
\providecommand{\pp}[1]{\left[#1\right]} 
\providecommand{\pr}[1]{\left(#1\right)} 
\def\bep{\begin{pmatrix}} \def\eep{\end{pmatrix}}
\def\bev{\begin{vmatrix}} \def\eev{\end{vmatrix}}

  \def\resp{respectively}

\def\m0{{\mathcal R}_0} \def\la{\label}
\def\Eq{\Leftrightarrow}\def\qu{\quad}
\newcommand{\bv}[1] {\boldsymbol{{#1}}}

\def\vi{\bv{i}}

\newcommand\s{\boldsymbol{s}}

\renewcommand{\epsilon}{\varepsilon}	
\def\w{\omega}

\newcommand\0{\mathbf{0}}

\newcommand*{\Scale}[2][4]{\scalebox{#1}{$#2$}}%
\DeclareMathOperator\ch{char} %

\newcommand{\inD}[1][\relax]{\def\argone{#1}\def\temprelax{\relax}
\ifx\argone\temprelax\right.\else\,\middle|#1\right.{}\fi}

\allowdisplaybreaks

\def\s{\;\mathsf s}\def\al{\alpha}
\def\a{\;\mathsf a}\def\wl{w.l.o.g. } 
\renewcommand{\i}{\;\mathsf i}
\renewcommand{\r}{\; \mathsf r}
\def\m{b}\def\sec{\section}\def\ME{mathematical epidemiology}

\def\mG{\mathcal G}\def\Dt{Descartes type}
\newcommand\red[1]{\textcolor{red}{#1}}

\def\m{\Lambda}\def\xd{x_{dfe}}\def\z{\;\mathsf z}\def\ch{characteristic polynomial }\def\pol{polynomial }
\newcommand{\figu}[3]{
\begin{figure}[H]
\centering
\includegraphics[scale=#3]{#1}
\caption{#2\label{f:#1}}
\end{figure}
}
\begin{document}
\title{Algorithmic approach for an unique definition of the next generation matrix
}
	\author{
 Florin Avram$^{1}$,  Rim Adenane$^{2}$, Lasko Basnarkov$^{3}$, Matthew Johnston$^{4}$ 
}
\maketitle

\begin{center}
	
		$^{1}$
		Laboratoire de Math\'{e}matiques Appliqu\'{e}es, Universit\'{e} de Pau, 64000, Pau,
 France; avramf3@gmail.com \\
$^{2}$ \quad Laboratoire des Equations aux dérivées partielles, Alg\'ebre et G\'{e}om\'{e}trie spectrales, d\'epartement des Math\'ematiques, Universit\'e Ibn-Tofail, 14000, Kenitra,
 Maroc; rim.adenane@uit.ac.ma \\
 $^{3}$ \quad  Faculty of Computer Science and Engineering, Ss. Cyril and Methodius University in Skopje, 1000 Skopje, North Macedonia; lasko.basnarkov@finki.ukim.mk\\
 $^{4}$ \quad Department of Mathematics + Computer Science
Lawrence Technological University, 21000 W 10 Mile Rd, Southfield, MI 48075, USA ;  mjohnsto1@ltu.edu
	\end{center}
	
	\maketitle

\begin{abstract}

  The basic reproduction number $R_0$ is  a concept which originated in population dynamics, \ME, and ecology, and is closely related to the mean number of children in branching processes (reflecting the fact that the phenomena of interest are well approximated
by branching processes, at their inception). Despite the very extensive literature around $R_0$ for deterministic epidemic models, we believe there are still aspects which are not fully understood. Foremost is the fact that $R_0$ is not a function of the original ODE model, unless  we include in it also a certain $(F,V)$ gradient decomposition, which is not unique. This is related to the specification of the "infected \com s", which is also  not unique.  A second interesting question is whether the extinction probabilities of the natural continuous time Markovian chain approximation of an ODE model around boundary points (\DFE\ and invasion points) are also related to the $(F,V)$ gradient decomposition, as suggested below in Section \ref{s:BD}.

We offer below three new contributions to the literature:

1) We offer  a universal algorithmic  definition of a $(F,V)$ gradient decomposition (and hence of the resulting $R_0$),  which requires a minimal input  from the user, namely the specification of an admissible  set of disease/infection variables.
We also present  examples where  other  choices may be more reasonable, with more terms in $F$, or more terms in $V$. This trade-off is explained briefly in section \ref{s:NGM}, Remark \ref{r:tra}, and further examined in  examples in Sections \ref{s:Lee}, \ref{s:KriOlaIRN}.

2) We glean out from the works of Bacaer a fixed point equation \eqr{qeq} for the extinction probabilities of a stochastic model  associated to a deterministic ODE model, which    may  be expressed in terms of the  $(F,V)$ decomposition. The fact that both $R_0$ and the extinction probabilities are functions of $(F,V)$ underlines the centrality of this pair, which may be viewed as more fundamental than the famous \NGM\ $F V^{-1}$. The equation \eqr{qeq} may be rarely solved explicitly; however, even when this is not the case, useful quasi-explicit solutions may be provided via "rational univariate representations" (for which an algorithm which works often is also provided).

3) We suggest introducing  a new concept of sufficient/minimal disease/infection set (sufficient for determining $R_0$). More precisely,
our universal recipe of  choosing "new infections" once the "infections" are specified suggests focusing on the choice of the latter, which is also not unique.
The  maximal choice of choosing all \com s which become $0$ at the given boundary point seems to always work, but is  the least useful for analytic computations, therefore we propose to investigate the minimal one.
As a bonus,   this idea seems useful for understanding the  Jacobian factorization approach for computing $R_0$ (see below).   We view   this as a method for obtaining an approximation, 
which we show to always yield  upper or lower bounds of the true $R_0$, depending on whether $R_0 \le 1$ or not. This raises  interesting questions of determining conditions on the epidemic model which ensure tightness of our bound,  of getting better bounds when tightness doesn't hold, and seems to deserve further work, only the surface of which is skimmed  below.

Last but not least, we offer Mathematica scripts and implement them for a large variety of examples, which illustrate that our recipe offers always reasonable results, but that sometimes other reasonable $(F,V)$ decompositions are available as well. 

\end{abstract}

 \tableofcontents

\section{Introduction}

{\bf Motivation}. Mathematical epidemiology has started by proposing  simple models for specific epidemics, and computing explicitly certain important characteristics like the \brn\ and the final size;   for example the SIR model was introduced, among other concepts, in the celebrated "A contribution
to the mathematical theory of epidemics" \cite{Ker}. The most fundamental and actually only general result of the field, due to Diekmann, Heesterbeek, Van den Driesche and Watmough, is expressing the  \DFE\ stability domain in terms of
$R_0$, which is defined as the \PF\ eigenvalue of a certain $(F,V)$  gradient decomposition (this is presented in detail in Section \ref{s:NGM}).
But, since the $(F,V)$   decomposition is not unique, it seems to us that the question of  what is $R_0$ still  deserves further  discussion.

On the other hand, one may note that nowadays, mathematical epidemiologists typically either restrict themselves  to low-dimensional models, resolved symbolically,  even by hand, or consider very complicated  models which are  resolved only numerically, for particular values gleaned from the medical literature. Missing from here are moderately complex models, which may be solved partly symbolically for any values of the parameters, but where the use of computer algebra systems (CAS) is either indispensable or greatly facilitating. Even in the case of papers belonging to this level -- see for example \cite{KriOla}, the role of the CAS is deemphasized.   Our paper is also an attempt to cast the CAS as  one of the main heroes of our story.

{\bf Our main result}. We provide below, for the first time, a  universal recipe for choosing a natural $(F,V)$  gradient decomposition, which only requires specifying the disease compartments (a subset of those which are  zero for the boundary point under consideration) (informally, these are not far conceptually from the so called fast components of singular perturbation theory). This decomposition is useful both for determining $R_0$ and for computing the extinction probabilities of an  associated  stochastic model.
We identify also examples in which the $(F,V)$ decomposition is not unique, and in which choosing another decomposition  with $F$ of lower rank may be beneficial for simplifying  the $R_0$ formula.

{\bf First restriction} (among others to follow). In this paper, we will restrict to   mathematical epidemiology models for which  there  exist  at least two possible special fixed states.
   The first, the  \DFE\ (DFE),  corresponds to the elimination of all possible compartments involving sickness, and will be assumed to be { unique}. Typically, this point is locally stable
   only for certain values of the parameters. Outside this domain  it is typically replaced  by another fixed point, which will be called ``endemic" if all its components are positive, and "resident boundary point" otherwise.

   Importantly, the stability of  the  DFE may be related to the historically famous {\bf basic
reproduction number}  and  {\bf net reproduction rate}--see \eqr{R0R}.
These  pillar concepts in  population dynamics,  mathematical epidemiology,  virology, ecology, etc, were  already introduced by the father
of mathematical demography Lotka  -- see \cite{lotka1939analyse,dietz1993estimation}, and also the introduction of the book \cite{bacaer2021mathematiques}.

{\bf  A bit of history of the net reproduction rate $\mR$, and its evolution into the mathematical concept  of basic reproduction number/stability threshold $R_0$}. Loosely speaking, in the case of only one infectious class, the net reproduction rate $\mR$ describes
 the expected number of secondary cases which {\bf one infected case} would produce in a  homogeneous,
{\bf completely susceptible}  population, during the lifetime of the infection. This description is especially
relevant at the start of an epidemics,  when the dynamics is well approximated by that of a branching process (a fact which goes back to  Bartlett and Kendall -- see for example \cite{Diek,kendall2020deterministic}). The main characteristic  of a  branching process is  the ``fertility", i.e. the expected number of descendants  one individual produces in the next generation. As a consequence,  the
branching result  insuring extinction when the  fertility is less than one translates in  epidemiology into
 local stability results  of the disease
free equilibrium involving $\mR$.

 The reproduction number  $\mR$ intervened already, in a particular case, in the  foundational paper ``A contribution to the mathematical
theory of epidemics" \cite{Ker}, which showed that:
\BEN \im  The  condition \be{R0R} R_0<1, \quad \text{where } R_0=\sd \mR,\ee implies local stability of the DFE. Here $\mR$ is  the net reproduction rate (number of secondary infections  produced by one infectious individual), and    $\sd$ is the fraction of susceptibles at the DFE.
\im The  condition $$R_0>1$$ implies instability of the DFE.
\EEN

With more infectious classes, one deals at the inception of an epidemics with approximate  multi-class branching processes, whose stability is determined by  a ``{\bf \ngm\ }" (NGM) --see section \ref{s:NGM}.

{\bf  The ``Jacobian approach"   for computing $R_0$}.
For big size problems, this approach is  doomed to fail symbolically, since it is equivalent to the Routh-Hurwicz conditions (RH),
which rarely succeed symbolically beyond dimension $4$ (also, RH is irrelevant numerically,  since the eigenvalues themselves are just as easy to compute). Therefore,  we studied below a variant,  the ``Jacobian factorization approach",  which focuses on  an approximation,  which we show to yield always upper or lower bounds of the NGM $R_0$, depending on whether $R_0 \le 1$ or not -- see Theorem 1. Several questions around  this bound are scattered below  in  Sections \ref{s:Bulh},  \ref{s:Feng}, \ref{s:KriOla}.

Note, as mentioned in \cite{heffernan2005perspectives},   that an example where the Jacobian
method does not yield $R_0$  is offered in Diekmann \& Heesterbeek (2000, Exe 5.43), and that
Roberts \& Heesterbeek (2003),  suggest that when
 threshold parameters determined from the Jacobian do not have the
biological interpretation of dominant eigenvalue
of the next generation matrix, then they should not be
called  basic reproductive ratios, nor denoted $R_0$ (we  follow their suggestion and use the notation  $R_J$ in this case).

{\bf The dilemma of the several different methods for computing $R_0$}  has  been  discussed in many papers, see for example
\cite{heffernan2005perspectives,li2011failure}. But this is a direct
consequence of the non-uniqueness of the $(F,V)$ decomposition.

 {\bf Deterministic or stochastic models?} Most of the mathematical epidemiology papers belong exclusively to one of these two paradigms.  However, any deterministic model may also be viewed as a stochastic continuous time Markov field (CTMC), evolving on the integers. One interesting CTMC, which seems not to have been discussed before, is presented in section \ref{s:BD}.

{\bf Contents}. Our paper is structured as follows. Section \ref{s:ME} recalls the definition of the DFE and provides our algorithmic definition of the (F,V) decomposition, in the form of a Mathematica script, as well as a discussion -- see Remark \ref{r:tra}-- of why other decompositions might turn out useful. This section also provides a new equation \eqr{qeq} for computing extinction probabilities for associated continuous time Markov chain models in terms of  the (F,V) decomposition, and shows that the Jacobian factorization approach yields upper bounds and lower bounds for NGM $R_0$'s, in section \ref{s:Jac}.

We turn then to a series of examples, chosen to help investigating what may be the major open problem in the field nowadays, which,  in our opinion,  is relating on one hand  $R_0$, and on the other hand  the extinction probabilities -- see below-- and the duration of minor epidemics
\cite{allen2012extinction,AllenVdD,Scoglio,tritch2018duration,nandi2019stochastic}, which is not further touched here.

Some hint that such a connection might exist may be gleaned from the SIR case studied by Whittle, and the SEIR  case, recalled in section \ref{s:SEIR}.

 Section \ref{s:Guo}  presents a host only model, with a single susceptible class and an $F$ matrix of rank one, where the  formula of $R_0$ may be ``guessed by inspection" of  the flow chart.
 This kind of examples have kept alive the hope of ``interpretable $R_0$ formulas", as illustrated in other recent papers -- see for example
 \cite{guo2022computing,segovia2022petri}. These authors  start by presenting simple cases, and  propose then algorithms for more complex cases based on the graph
 structure of the flow chart, which in our opinion are not sufficiently detailed or documented. While it may well be that tools like Petri nets, as proposed in the second paper, will one day  succeed for resolving flow chart with certain structures, this does not seem
 to have happened yet. Also, for models with \NGM\ of high-rank,  the lack of simple formulas for $R_0$ and of "simple biological interpretations" is natural to be expected; simple formulas for the spectral radius can only be a consequence of a simple graph structure which has not been pinpointed yet.

Sections \ref{s:Lee}, \ref{s:Bro22}, offer
 two examples in which several $R_0$ formulas were offered in the literature, but  we are at a loss of how to choose among them. In the first case (a virus-tumor model), the recipe $R_0$ is simpler than its competitor, but in the second case (a vector-host model), it is more complicated.

Section \ref{s:inv}  shows that the boundary equilibria, and the (invasion) reproduction numbers may be easily computed with our scripts; to illustrate this, we use a two-strain host only   model from \cite[Ch.8]{Mart}, where our recipe  NGM yields    the same answer as that   given by the Jacobian factorization.

 Section \ref{s:Bulh} offers another two-strain host only example, this time including also vaccination,  in which our recipe  NGM yields again   the same answer as that   given by the Jacobian factorization.


 Section     \ref{s:Mart15} offers an example from the textbook of \cite{Mart} in which the square relation stops holding.

 Sections        \ref{s:Feng}, \ref{s:KriOla} offer yet more examples, this time in the two-strain vector-host context, in which our recipe  NGM yields   an $R_0$ formula which is precisely the square root of that given by the Jacobian approach. Note that here the first of the
 {three  elegant relations concerning the invasion numbers from Section \ref{s:Bulh} --  see Remark \ref{r:mys}--holds, but the other two seem to break
 down.}

The last subsection provides, for the invasion numbers,  a second   example where  another choice of $R_0$ may be more reasonable,  on the grounds of leading to a simpler answer (but the admissibility requirement forces  then extra assumptions on the parameters).

\sec{A bird's eye view of \ME: the \DFE, the \NGM, and an algorithmic definition  of a stability threshold associated to the \brn\ $R_0$}\la{s:ME}

\ssec{The disease free equilibrium (DFE)}
  The DFE may be defined as a ``maximal boundary state", and    may be  found by identifying   a maximal sub-system of the ODE epidemic model which factors
  \be{Kfac}\vi'= M \vi,\ee
  where prime denotes derivative with respect to time, and $M$ is a  matrix may depend on $\vi ,$ but may not explode in the domain of interest, which we will take for simplicity to be $\R_+^n$.

\beR One fixed point of this system is $\vi=0$.  This motivates us to call the components  $\vi$  disease or infectious states. The set of all its  indices will be denoted by $\mI$.  Note that specifying `$\mI$  induces a partition of both the coordinates and the equations of our original system into infection (eliminable) components, and ``non-infection" (the others).

 The eventual other fixed  points may be found by solving $M=0$ together with the other non-infection equations under the condition $\vi=0$.\eeR

        In this paper we will assume   uniqueness of the DFE, at least after excluding   biologically irrelevant fixed points, like an unreachable origin.

  We end this section with the    very elementary script that implements this.  Note that any ODE model ``mod" (like SIR, etc...), is   a pair mod= (dyn,X) consisting of a vector field ``dyn" and a list of variables ``X", and that to find any boundary fixed point it suffices to know the set of indices ``inf" where it is $0$, so that we solve the system ``dyn==0" under the condition ``X[[inf]]->0". But, since sometimes only numeric solutions are possible, our DFE Mathematica script below has also an optional numerical condition parameter ``cn", which is taken by default as the empty set.

  \begin{verbatim}
 DFE[mod_,inf_,cn_:{}]:=Module[{dyn,X},
  dyn=mod[[1]]/.cn;X=mod[[2]];
  Solve[Thread[dyn==0]/.Thread[X[[inf]]->0],X]];
 \end{verbatim}
For the non-Mathematica users,  only the Solve command is relevant, the others being just Mathematica implementation details.

\subsection{$(F,V)$ gradient decompositions, the next generation matrix, $R_0$, and a simple  recipe for computing them}\la{s:NGM}
From now on, the infection equations \eqr{Kfac} will be rewritten as
\be{dec}\vi'=\mF-\mV= (F-V) \vi. \ee

Of course, such a decomposition is not unique, but we will also ask, following \cite{Diek,Van,Van08}, that $F$, the gradient of $\mF$, is a  matrix with nonnegative elements, and $-V$, the gradient of $-\mV$ is a Markovian generating matrix (i.e. a matrix with non-negative off-diagonal elements and non-positive row-sums). Conceptually, $F$ models inputs to the disease compartments from outside ("new infections"), and $-V$ models transfers between  the disease compartments. Still, a priori, the decomposition \eqr{dec}
is not unique.

\beXa Let us illustrate this via a SIR example with superinfection, in which the classes S and R play symmetric roles, inspired by the works of  \cite{Mogh,Jin,NillI}
\bea
\bc\s'(t) &=
\L_s -    \s(t)   \pp{ \b_s  \i(t)(1+ \xi  \i(t))+d_s}+i_s   \i(t)+ \g_r \r(t)  \\
\i'(t) &=
 \i(t)\Big[\pp{\b_s \s(t)+\b_r \r(t)}(1+ \xi  \i(t))\Big]  - d_i \i(t) \\
\r'(t) &= \L_r -\r(t)\pp{ \b_r \i(t)(1+ \xi  \i(t))+d_r } +i_r   \i(t)+ \g_s \s(t)\ec.
\eea
Here, the only infection equation, the second, is already written in a decomposed form
$\mF-\mV, \mV=d_i  \i(t)$,
and $
 F=\Big[\pp{\b_s \s(t)+\b_r \r(t)}(1+ \xi  \i(t)) \Big]+ \xi \i(t) \pp{\b_s \s(t)+\b_r \r(t)}$.

 Note that for the application of the \NGM\ method we must plug finally $ \i=0$;  therefore, the second term in $F$, due to "superinfection",  is  irrelevant for this purpose.
 \eeXa

\beR The possible non-uniqueness of the decomposition brings us to a delicate point in \ME. Anticipating a bit, since $R_0$ is the \PF\ eigenvalue of $F V^{-1},$ strictly speaking, $R_0$ is not determined just by an ODE epidemical model, but also by the $(F,V)$ gradient decomposition.  If we want that  an ODE epidemical model determines uniquely
an $R_0$, we must include in the definition of an ODE epidemical model also the $(F,V)$ gradient decomposition we adopt.

\eeR

\beD \la{d:EM}  An ODE epidemic model is an ODE  dynamical model in which a certain subset of  equations, usually called "disease/infection" equations, and to be referred from now on also as {\bf zeroable set},     admits at least one admissible decomposition \eqr{dec}, with $(\mF, \mV)$ satisfying the conditions (A1-A5) of \cite{Van08}.

\eeD

\beR
 Note that \eqr{dec} is the most common model used in population dynamics.  This makes natural
 to define informally ODE epidemic models as population dynamics models \eqr{dec},  with extra equations modeling  interactions with the non-disease \com s, which   admits at least one admissible decomposition. \eeR

\beR \la{r:def} The definition of ODE epidemic models above is  imprecise, since it does not list all the requirements we must put on an ODE model. Some reasonable    restrictions are \BEN \im Essentially nonnegative processes having  a non-empty set of disease classes, so we deal with an epidemic
 (note however that we define disease classes in the sense of classes which satisfy \eqr{Kfac}, which excludes for example importation models). \im
 Processes  with a unique DFE,  at least after excluding   biologically irrelevant fixed points, like an unreachable origin.
 \im   The  local stability domain  of the DFE is non-empty, and not the full set.
 \im The dynamical system has polynomial coefficients, to be able to take advantage of the remarkable symbolic computation tools available
 for this class.
 \EEN
   We make these assumptions because they are satisfied by most mathematical models which have already been used for modelling real life biological phenomena.  However, these  assumptions might not be enough, and   further ones   might be necessary  for obtaining the currently missing precise definition of ``real life ODE  \ME\ models".\eeR

\beR \la{r:tra}
Admissible decompositions need not be unique.  A priori, one may "move terms from F to V", to lower its rank and simplify the formula for $R_0$, and also "move off-diagonal terms from V to F", which enlarges the domain of parameters which ensure that $V^{-1}$ has positive terms. There is  a tradeoff between these two possible moves, since simplicity  of  $R_0$ comes at the cost of extra assumptions on the parameters.  Our universal   decomposition seems to strike a balance between the two directions.
\eeR

   \beR It was  emphasized from the outstart -- see for example  \cite{diekmann2010construction,li2011failure,cushing2016many,Van17}, that  an ODE mathematical epidemiology model might have several ``admissible decompositions", which might yield  distinct next generation matrices and distinct $R_0$'s.

 This was   viewed as a richness rather than a default.  A similar point of view is  advocated in the recent paper \cite{Lee}, which argues that the class of all admissible NGM's associated to an ODE epidemiology model must be studied as an ensemble, rather than focusing on individual representatives.

\eeR

For any admissible  decomposition,  Diekmann, Heesterbeek,  Van den Driesche and Watmough established  the following celebrated  DFE stability theorem:

\beP \la{L1} For any admissible  decomposition $(F ,V)$, let $$R_0=\rho(F V^{-1})$$ denote the \PF\ eigenvalue  of the {\bf  next generation matrix}.
Then, the  DFE is unstable on  $R_0>1,$ and
  locally  stable on $R_0< 1$   \cite{Diek,diekmann2000mathematical,Van, Van08}.
  \eeP

For a recent historical overview of  $R_0$, next generating matrices,  and their calculation in many examples, we refer  the reader to the delightful paper \cite{brouwer2022spectral}.

 Unfortunately, the standard definition of a next generation matrix (and hence of $R_0$)  involves  concepts like  ``new infections", which  were defined in the original papers  based on  epidemiological  considerations,   and require therefore the intervention of a human expert. This had created the impression that this method can not   be
encapsulated into a computer program. However, we offer and implement below a  simple algorithmic definition, based only  on the structure of the system and of the ``infectious/disease equations".

Our proposal is to use a special F-V decomposition with F constructed as the positive part of all the interactions in the disease equations which involve both disease \com s and input/susceptible ones. The latter are defined as the complement of the disease \com s, after the possible removal of output \com s, which may be specified as deterministic functions of the other \com s (i.e. may be computed, once the other \com s are known). Note now that the concept of "positive part of  the interactions" may be hard to pinpoint mathematically, but useful enough to have been implemented in CAS's (Mathematica, Maple, Sage, etc); this made us adopt the following definition:

\beD For a given zeroable set,  an admissible $(F, V)$  gradient
decomposition \eqr{dec} is one where $F$, the gradient of  $\mF$, does not contain in its expanded form syntactic minuses in its CAS representation,  and where $V$,  the gradient of  $\mV$ is such that $-V$ is a ``sub-generating matrix", under the assumption of nonnegativity of all the model parameters. Note that this last condition implies, but is not equivalent to  the existence of $V^{-1}$ and to the nonnegativity of all its elements.
\eeD

The problem of whether the $R_0$ of the decomposition provided  satisfies under certain conditions the   stability theorem of Van den Driessche and  Watmough  is still open; therefore it should be  viewed for now just as a recipe that  works well in simple cases.

After lots of experimenting, we have found only few cases -- see for example Section \ref{s:KriOla} where the recipe NGM has
a serious competitor; it is for computing the invasion reproduction number for a two-strain vector-host model with altered infectivity for co-infected vectors,  and with ADE (antibody-dependent enhancement).

\ssec{An algorithmic $ F-V$ decomposition}\la{s:FV}
We complement now the famous $\mF-\mV$ ``equations decomposition" and \NGM\ method of \cite{Diek,Van,Van08}  by an algorithmic $ F-V$ decomposition.
\BEN \im
The user supplies the model ``mod" (a pair containing the RHS of the dynamical system, and its variables), and  the indices ``inf" of the disease (or infectious) variables; the indices of the
other compartments are denoted by ``infc".
\im Subsequently, the Jacobian of the infectious equations $M$ with respect to the corresponding variables is computed.
\im Define the interaction terms   as terms in $M$ which contain variables $s\in \text{infc}$, and which, if positive, must end up in $F$.
Their complement, denoted by $V1$,  will form part of $V$.
\im A first guess  for  $F$, $F1$ is constructed as the complement of $V1$.
It contains all the interaction  terms (which involve  both disease and susceptible \com s).
\im  $F$ is obtained by  retaining  only the positive part of the matrix $F1$, i.e. the terms which do not contain syntactic minuses.\fn[4]{we use the simplest algebraic representation of the equations, and do not study the effect which algebraic manipulations introducing minuses might have.} Finally, $V1$ is increased to $V$, which is the complement of $F$.
\im The script outputs \{M,V1,F1,F,V,K\}.
\EEN

 \begin{verbatim}
NGM[mod_,inf_]:=Module[{dyn,X,infc,M,V,F,F1,V1,K},
   dyn=mod[[1]];X=mod[[2]];
   infc=Complement[Range[Length[X]],inf];
   M=Grad[dyn[[inf]],X[[inf]]]
   (*The jacobian of the infectious equations*);
   V1=-M/.Thread[X[[infc]]->0]
   (*V1 is a first guess for V, retains all gradient terms which
   disappear when the non infectious components are null*);
   F1=M+V1
   (*F1 is a first guess for F, containing all other gradient terms*);
   F=Replace[F1, _. _?Negative -> 0, {2}];
   (*all terms in F1 containing minuses are set to 0*);
   V=F-M;
   K=(F . Inverse[V])/.Thread[X[[inf]]->0]//FullSimplify;
{M,V1,F1,F,V,K}]
 \end{verbatim}

Note that  our NGM script requires a minimal  input from the user, just the specification of the disease \com s; there is no need to specify ``new infections".

  The results of this decomposition   seem to yield correct results in all the examples from the literature we checked. We would like to add   that for dynamical systems satisfying the four conditions in the remark \ref{r:def}, this decomposition yields ``admissible gradient decompositions", in the sense that $V^{-1}$ will contain only non-negative terms, and that it is furthermore obtainable from an equations decomposition which is admissible in the sense of \cite{Van08} (see Definition 1), and yields therefore
  the correct stability domain.

  \beR Note that the ``Replace" command in the script uses the powerful Mathematica capability of applying a ``rule" to  parts of an ``expression", specified by ``levelspec",  and that it was furnished to us by the user Michael E2 in \begin{verbatim}
https://mathematica.stackexchange.com/questions/286500/
how-to-set-to-0-all-terms-in-a-matrix-which-contain-a-minus
/287406?noredirect=1#comment715559_287406
\end{verbatim}

\eeR

Finally, let us discuss an alternative possible implementation. We could  just provide NGM with the right-hand side of the differential equations, compute the steady states, specify one of them, and then define the infected classes as the components with zeros.

However, this would be unpractical,  since for the majority of the models with explicit DFE the other fixed points are either not explicit, or require very long execution times. It is therefore much simpler to have the user help the AI by providing it with  $\mI$, which leads immediately to the matrix $M$.  Essentially, we jump directly to the factorization \eqr{Kfac} of the infected equations,  postponing solving   the non-infection variables later. 
\ssec{A multi-dimensional birth and death CTMC process associated to a $(F,V)$ decomposition, its branching process approximation, and the Bacaer equation for the probability of extinction}\la{s:BD}

The works of Kendall and Bartlett suggest that   ODE epidemic models may be associated to corresponding birth-and-death CTMC processes, and then approximated further by branching process.

Citing \cite{Grif} :"It has been noted by Bartlett (1955), p. 129, that for an epidemic in a large
population, the number of susceptibles may, at least in the early stages of an
outbreak, be regarded as approximately constant at its initial value and that this
approximation will continue to hold throughout the course of an epidemic,
provided that the final epidemic size is small relative to the total susceptible
population. Thus the general epidemic process may be approximated by a simple
birth-and-death process."

To make this more precise, a $(F,V)$ decomposition \eqr{dec} determines a natural associated multi-dimensional birth and death CTMC process, by fixing the values of the  non-disease variables, so that the matrices $(F,V)$ depend only on $\vi$, and interpreting the transition rates between compartments as rates of BD transitions.
If the CTMC has rates which are linear in the disease variables, one may associate to it a branching process, and take advantage of the well-known equation for extinction probabilities.    This procedure has been detailed in previous works like
\cite{allen2012extinction,AllenVdD,Scoglio,tritch2018duration,nandi2019stochastic}, and used to approximate  extinction and invasion probabilities,  as well as the  duration of minor epidemics. If the CTMC has rates which are super-linear in the disease variables, a further approximation of ignoring the higher power terms in $\vi$ is necessary. At the end, this results in assuming that the matrices $(F,V)$ are constant (do not depend  on $\vi$).

Let us illustrate this philosophy on the famous SIR example. However, in line with our interest in this paper and getting a bit ahead of ourselves, we will only look at an  "disease process" of the infected, with the other components fixed. The state space of the process will thus be $\N$. We note this is similar in spirit with the "slow-fast/singular perturbation" technique of considering only variables whose  lifetime is short, and fixing the other variables whose  lifetime is longer, and is  in fact the idea behind the famous \NGM\ approach.

 \beXa The "SIR" disease process (i.e. defined on the disease \com s) is $i'=(\b s-\g)i$.
  The natural {\bf SIR/linear CT birth and death} disease stochastic process (DSP)  is a Markov process $X_t \in \N$
  with  generating operator on the set of  functions $f:\N ->\N$ defined by
\be{gen} \mG f(i)= \b s i (f(i+1)-f(i)) + \g i( f(i-1)-f(i))=A f(i),\ee
{and  corresponding semi-infinite generator matrix
\be{gmat} A=\bep -\beta & \beta &0&\cdots& 0\\\gamma&-\beta-\gamma & \beta&\cdots&0\\0&\gamma&-\beta-\gamma&\beta&\ddots \\
\vdots & \ddots& \ddots&\ddots & \beta\\
0&0 & \cdots& \gamma&-\beta-\gamma
 \eep.\ee}

\beR We recall for the benefit of readers who have not been exposed to the (immense) literature on Markov processes that  the  behavior of expectations of this class of   stochastic processes always involves one deterministic operator $A$, the generator of the Markovian semigroup, which acts on a space of "appropriate functions" on the state space \eqr{gen}, and where  "appropriate" may be skipped in simple cases like ours  \eqr{gmat}. The essential thing to notice here is that our Markov generator operator
$A$ is completely defined by the rates, just like its "mean-field" deterministic ODE.
Thus, from the practical point of view of estimating rates, we have added nothing to  the parameters of the ODE model (as would be the case with other stochastic processes involving Brownian motion, etc). We have only modified the state space and the operator; however, this way, phenomena which are invisible in the continuous mean-field limit, become relevant.

Finally, for readers puzzled by the question of where is the randomness hidden in the
deterministic operator  \eqr{gen}, we mention that this arrives via two Poisson processes describing the times when the process jumps up and down, \resp, and we refer to the literature for more details. \eeR

This process converges to $\I $ (i.e. is non-recurrent) or to a stationary distribution  iff
$R_0:=\fr {\b s}\g$ is strictly bigger than $1$, or strictly smaller
than $1$, respectively. The probability of "extinction/absorbtion into $0$",
when starting  the process with  $j$ infected are
\be{prS}p(j)=q^j, \qu q=\bc 1& R_0 < 1\\ \fr \g{\b s}=\fr 1{R_0} & R_0 \ge 1 \ec.\ee

This result may be found for example in the textbook \cite{dawson2017introductory}
(it  is,
up to technical difficulties caused by the non-compact state space,
the simplest  illustration of the fact that solutions of "Dirichlet problems"
of the form $p(j)=E_{X_0=j}[g(X_{\tau})],$ where $\tau $ is the exit time from a
domain $S$, must solve $\mG p=0$  and  $p=g$ on the boundary of $S$).

The expected time to extinction when starting  the process with is $j$ infected and when $R_0 <1$ may be found using the fact that solutions of "Poisson problems" of the form $T(j)=E_{X_0=j}[\int_0^{\tau}h(X(s) ds],$  must solve
$$\bc \mG T + h=0\\  T=0 \text{  on the boundary of } S\ec.$$

Another interesting quantity is the expected time to extinction when $R_0 >1,$ in the case that extinction occurs. This "Dirichlet-Poisson  problem" may be written as
 $$T(j)=E_{X_0=j}[g(X_{\tau}) \int_0^{\tau}h(X(s) ds],$$ where $h=1$, and $g$ is the indicator of extinction occurring.  Such expectations must solve
 $$\bc \mG T + h p=0\\  T=0 \text{  on the boundary of } S\ec,$$
 where $p$ is the solution of the Dirichlet problem with boundary value $g$.

For SIR, we must solve respectively
 \be{recS}\Scale[0.86]{ \bc \b s x (T(x+1)-T(x)) + \g x( T(x-1)-T(x)) + 1=0, T(0)=0, T(K)=0, K->\I,&\text{when  } R_0 <1\\ \b s x (T(x+1)-T(x)) + \g x( T(x-1)-T(x)) + q^x=0, T(0)=0, T(K)=0, K->\I,& \text{when  } R_0 \ge 1\ec.} \ee

These two equations may be solved explicitly. The limits are quite challenging even with Mathematica, as shown in the Appendix \ref{s:Whi}. We are able to recover and generalize the results of \cite{whittle1955outcome} (see also \cite[(10)]{tritch2018duration})  when $j\ge 1$ for the first problem, but not for the second one.

\eeXa

{\bf The Bacaer equation}. One missing aspect in the previous works however is characterizing the extinction probabilities via one final equation, without going through the discretization procedure employed in \cite{allen2012extinction,AllenVdD,Scoglio,tritch2018duration,nandi2019stochastic}, and solving each example individually.
Interestingly,  such an equation in terms of $(F,V)$ decompositions
  was provided by Griffiths in \cite{Grif}, except that this paper  considers only BD's with no transfers.

We review now the work of \cite{BacAit} (who were motivated
  by  analyzing the case of periodic steady solutions), but on the way spelled out also the simple equation \eqr{qeq} below.
To each fixed values for the disease variables, one may  associate to a $(F,V)$ decomposition $\vi'= (F-V) \vi$  a "multi-dimensional birth and death process" (BD), with birth rates
 given by {$F $}, and with transfer and death rates given by  $-V$.\fn[4]{$\vi'= (F-V) \vi$ are precisely the mean-field equation for the
multi-dimensional birth and death process; this is precisely  \cite[(6)]{Grif}, under the extra condition that we assume, that the immigration vector into the disease \com s is $0$.} In fact,
 the $-V$ matrix by itself generates an  absorbing CTMC (and the $F$  matrix models roughly inputs to be fed into this absorbing CTMC). This observation explains
 that an  ODE \ME\ model has associated to it both a birth and death process, as well as a "death and transfer only" absorbtion CTMC
 --see Remark \ref{r:VFKr} for an example. If furthermore $(F,V)$ are independent of $\vi,$ we  are dealing with a branching process (approximation).

A useful fact to recall is that the probabilities of extinction of a multi-variate discrete time branching process  are of the form
$$\mathbb{P}_0=q_1^{j_1}q_2^{j_2}...,$$
where $q=(q_j, j=1,..., J)$, and $J$ is the number of disease \com s, and where $q_j$ satisfy the "Bacaer equation"
\be{qeq} (q^t \circ F) * q + (1-q)\circ V-q * f=0 \Eq q_j=\fr{\sum_{k=1}^J (1-q_k) V_{kj}}{f_j-\sum_{k=1}^J q_k F_{kj}},\ee
where $*$ denotes coordinate-wise product,  dot product is   denoted by $\circ$, and $f_j=\sum_k F_{k,j}$.
This equation is new, but it may be inferred from \cite[(9)]{Grif},  \cite[(11)]{BacAit}  and \cite[(5.3)]{BacR} (after some changes of variables).

For the SIR process for example, \eqr{qeq} becomes
$$(q-1)\b s q + (1-q)\g=(q-1)(\b s q -\g)=0,$$
with the two roots $q=1$ and $q=\fr \g{\b s}=\fr 1{R_0}$,  recovering Whittle's result (the two roots  yield the correct result when $R_0$ is strictly smaller
than $1$ and strictly bigger than $1$, respectively).

 We will check below that \eqr{qeq} also recovers  other explicit particular cases offered in the literature, like
SEIR  \cite{allen2012extinction}, \cite[sec.4]{AllenVdD}, SIV \cite{Milliken16},
etc

\ssec{The Jacobian factorization  bound}\label{s:Jac}
 Note first the following elementary fact:
 \beL A sufficient (but not necessary) condition for a polynomial with real coefficients and {\bf positive leading term}  to admit a positive
 root is that $c_0  <0$, where  $c_0 $
is the constant term   of the polynomial.

\eeL
For polynomials of degree $1$, this condition is  also necessary.
This converse result may be strenghtened  to ``\Dt\ polynomials".
\beD  \label{d:Dt} We will say that  a parametric polynomial with real coefficients, whose constant coefficient may change sign, but whose all  other coefficients are ``sign definite", and of the same sign (which may be supposed \wl\ to be $+$), is of \Dt.

\eeD
As an immediate consequence of Descartes's rule of signs, it follows that
\beL A sufficient and necessary condition for a Descartes polynomial with a {\bf positive leading term}  to admit a positive
 root is that $c_0  <0$, where  $c_0 $
is the constant term   of the polynomial.

\eeL

\beR Note the immense simplification with respect to the Routh-Hurwitz conditions,
when we need to establish the existence of a positive root, for a \Dt\ \pol.

We believe that ``the mystery of the success of the Jacobian factorization approach" comes from the fact that ``simple epidemic models" often feature \Dt\ polynomials. However, this leaves us with many further questions, like when does this happen and what to do when it doesn't.
\eeR

The  Jacobian factorization approach consists in:
\BEN \im Putting all the rational factors of the characteristic polynomial of the Jacobian, in a form normalized to have positive leading term, assuming they  are sign definite (if this is not the case, this approach does not work, but may be generalized).
\im Removing all linear factors  with  eigenvalues which are negative. 
\im For all remaining factors $F_i$ for which $c_0^{(i)} <0$ may hold for certain parameter values, rewrite this inequality into the form
$$c_0^{(i)} =c_+ - c_-=c_+(1-R_J^{(i)}) <0 \Eq R_J^{(i)}:=\fr{c_-}{c_+}>1,$$ where $c_+,c_-$ are the positive and negative parts of the expanded form of $c_0^{(i)}.$
\im Define the "Jacobian factorization $R_0$"
\be{RJ} R_J=\max_i[R_J^{(i)}].\ee
\EEN


\beT  A) In the instability domain, $R_J$ is a lower bound for $\inf_{F \text{ admissible}} R_F$.

 B) In the stability domain, $R_J$ is an upper bound for $\sup_{F \text{ admissible}} R_F$.

\eeT

{\Prf}   A) Fix any  admissible $F$ and let $R_F$ be its associated NGM $R_0$. Then
 $$R_J >1 \Eq \exists i: R_J^{(i)} >1 \Eq \exists i: c_0^{(i)} <0 \Lra \text{ DFE instability } \Eq R_F >1.$$  Thus
\be{Rin}R_J >1  \Lra  R_F >1 \Eq R_J \leq R_F,\ee
and the result follows.

B) Similar proof.


 {\bf Conjecture:} We conjecture that if all the  factors $F_i$ are Descartes polynomials, then $R_{J}=R_F$ for any admissible decomposition $(F,V)$, and will denote the resulting object by $R_0$.

   {\bf Open question 1:}  Under what
conditions do our NGM $R_0$ and our Jacobian  $R_J$  coincide?

The implementation  of the  Jacobian factorization approach is provided in Section \ref{s:Jac}.

\ssec{The ``rational univariate representation" (RUR) and the  reduced order quasi -stationary approximation}\label{s:RUR}

Hundreds of mathematical epidemiology
papers have already employed the idea of  reducing the fixed point system to one scalar equation in one of the disease variables, via rational substitutions for the other variables. We note  that this is a   particular case of the so-called "rational univariate representation" (RUR), but for Mathematica users this is irrelevant, since RUR is not  implemented currently, and we had to write our own script, included below, in which the user chooses in a system the variable he wants to restrict to.

The current code for this reduction to one equation algorithm is
\begin{verbatim}
RUR[mod_, ind_, cn_ : {}] := Module[{dyn, X, par, eq, elim},
      dyn = mod[[1]]; X = mod[[2]]; par = mod[[3]];
      elim = Complement[Range[Length[X]],ind];
      eq = Thread[dyn == 0];
      ratsub = Solve[Drop[eq, ind], X[[elim]]][[1]];
      pol =
       Collect[GroebnerBasis[Numerator[Together[dyn /. cn]],
         Join[par, X[[ind]]], X[[elim]]], X[[ind]]];
          {ratsub, pol}
    ]
\end{verbatim}
\beR The command which does the essential work is ''GroebnerBasis". When ''ind" is a set with just one  component, this reduces the system to a polynomial in this variable.
Alternatively, this could be achieved by plugging the results of  ''ratsub" into the system.

 The script above  works directly for models with demographics, but must be modified  for ``conservation systems", where the fixed points are only determined by adding the total mass conservation equation to the fixed point equations. \eeR

This script may also be used for order reduction, in the spirit of the  ( quasi-steady-state assumption) QSSA method in biochemistry, and of the recent epidemiology paper \cite{johnston2023two}. We illustrate this for the simplest SIR example.
\beXa For the SIR process $(S(t),I(t), R(t), t \geq 0)$ with linear birth rates $b_s,b_r$ for susceptible and recovered,  the system for the  fractions $\s(t)=\fr{S(t)}N, \i(t)=\fr{I(t)}N, \r(t)=\fr{R(t)}N, N=S+I+R$ is:
 \be{SIR}\bc
\s'(t)= b_s  -\beta \s(t) \i(t) +\g_r \r(t)- d_s \s(t),& d_s=\g_s + \mu\\
\i'(t)= \beta  \s(t)\i(t)-d_i \i(t),& d_i=\g_i+\mu +\de
\\
\r'(t)=b_r+  \g_i \i(t)+ \g_s \s(t)-d_r \r(t),& d_r=\g_r +\mu
\ec
\ee

The DFE is: $(\frac{b_r \gamma _r+b_s \left(\mu +\gamma _r\right)}{\mu  \left(\mu +\gamma _r+\gamma _s\right)},0,\frac{b_r \left(\mu +\gamma _s\right)+b_s \gamma _s}{\mu  \left(\mu +\gamma _r+\gamma _s\right)})$.

The rational substitution with respect to  $i$ obtained via RUR is:
$$\pr{s\to \frac{\gamma _r \left(b_r+b_s+i \gamma _i\right)+\mu  b_s}{\beta  i \left(\mu +\gamma _r\right)+\mu  \left(\mu +\gamma _r+\gamma _s\right)},r\to \frac{b_r \left(\beta  i+\mu +\gamma _s\right)+b_s \gamma _s+i \gamma _i \left(\beta  i+\mu +\gamma _s\right)}{\beta  i \left(\mu +\gamma _r\right)+\mu  \left(\mu +\gamma _r+\gamma _s\right)}.}$$
Note this reduces to the DFE when $i=0$.

The reduced approximate model obtained via RUR is:
\bea && i'=i \pp{a_0 - a_1 i}, a_1=\beta  \left(\mu  \gamma _i+(\delta +\mu ) \left(\mu +\gamma _r\right)\right),\\
&& a_0=\beta  \left(b_r \gamma _r+  b_s \left(\mu +\gamma _r\right)\right)-\mu \left(\mu +\gamma _r+\gamma _s\right)  \left(\delta +\gamma _i+\mu \right) \\&&
=\mu \left(\mu +\gamma _r+\gamma _s\right)  \left(\delta +\gamma _i+\mu \right) \pr{\sd \mR-1}=\mu \left(\mu +\gamma _r+\gamma _s\right)  \left(\delta +\gamma _i+\mu \right) \pr{ R_0-1}.\eea

This has an explicit (rather formidable) analytic solution, provided in the Mathematica file.

\figu{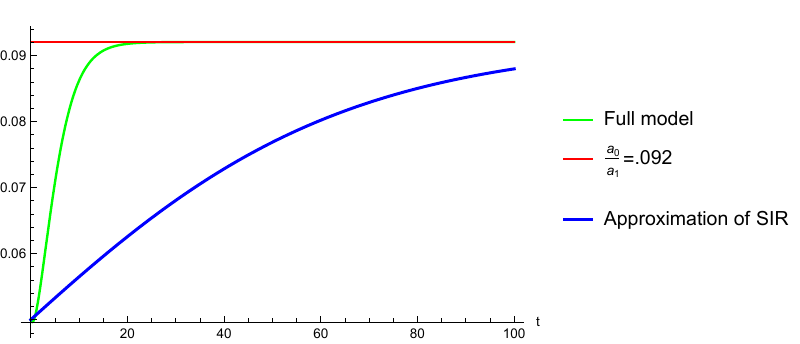}{{Illustration of the asymptotic convergence of  $i(t)$ towards the endemic value $\fr {a_0}{a_1}=.092$, both for the full SIR model and its approximation.}}{.7}

One may notice that for the chosen numerical illustration, the plots of $i$ and its approximation converge towards  the same value, but differ sharply for the chosen numeric values as far as  shape is concerned.
\eeXa

We mention finally the possibility to develop yet another possible algorithm for  computing a ``bifurcation $R_0$", suggested by the example above, which is based on the known fact that this parameter is expected to produce bifurcations  at $R_0=1$.

The steps are: \BEN \im Factor out the variable in the  scalar polynomial of the reduced model   (always possible iff this is  a disease variable).
\im  Write  the free  coefficient of the divided polynomial   as $F(R_0)=G(R_0) (R_0-1)$, where  $F(R)$ is rational (always possible due to the known bifurcation at $R_0=1$).
\im Identify  a factor which is linear in susceptible variables like $\sd$, etc,  and write it as a difference  of positive and negative terms.
 Upon  normalizing one of them to one, the other will be $R_0$, or $1/R_0$.
 \EEN

\section{$R_0$ and extinction probabilities for the   SEIR epidemic  model
\label{s:SEIR}}

The SEIR process $(S(t),E(t),I(t), R(t), t \geq 0)$ adds to the SIR model the class $E$(exposed). The model for the  fractions $\s(t)=\fr{S(t)}N,\e(t)=\fr{E(t)}N, \i(t)=\fr{I(t)}N, \r(t)=\fr{R(t)}N, N=S+E+I+R$ is:
 \be{SEIR}\bc
\s'(t)= b_s  -\beta \s(t) \i(t) +\g_r \r(t)- d_s \s(t),& d_s=\g_s + \mu\\
\e'(t)= \beta  \s(t)\i(t) - \g_e \e(t ) - d_e \e(t),& d_e=\g_e+\mu \no\\
\i'(t)= \g_e \e(t )-d_i \i(t),& d_i=\g_i+\mu +\de
\\
\r'(t)=b_r+  \g_i \i(t)+ \g_s \s(t)-d_r \r(t),& d_r=\g_r +\mu
\ec
\ee

This is both a textbook model, and one for which answers to many open questions (concerning for example the emergence of chaos under stochastic and periodic perturbations)  are still awaited  -- see \fe\ \cite{dietz1976incidence,schwartz1983infinite,forgoston2009accurate}.

The DFE of \eqr{SEIR} is: $(\frac{b_r \gamma _r+b_s \left(\mu +\gamma _r\right)}{\mu  \left(\mu +\gamma _r+\gamma _s\right)},0,0,\frac{b_r \left(\mu +\gamma _s\right)+b_s \gamma _s}{\mu  \left(\mu +\gamma _r+\gamma _s\right)})$. The decomposition matrices and \brn\ are:
$$ F= \left(
\begin{array}{cc}
 0 & \beta  s \\
 0 & 0 \\
\end{array}
\right), V= \left(
\begin{array}{cc}
 \gamma _e+\mu  & 0 \\
 -\gamma _e & \delta +\gamma _i+\mu  \\
\end{array}
\right), R_0=\frac{\beta  s \gamma _e}{\left(\gamma _e+\mu \right) \left(\delta +\gamma _i+\mu \right)}.$$
The associated disease stochastic process  ${X=(e,i)} \in \N^2$
  has generating operator

$$\Scale[0.9]{\mG= \b s i (f(x+e_1)-f(x)) + \g_e  e( f(x+tr)-f(x))+ d_e  e( f(x+e_3)-f(x))+ d_i  i( f(x+e_2)-f(x))},$$
where $x=(e,i), e_1=(1,0)=-e_3,  tr=(-1,1), e_2=(0,-1)$.

The extinction probabilities obtained by solving \eqr{qeq}  are
$$\bc q_i=1,q_e=1 &  \text{when  }R_0<1\\q_i= \fr 1 {R_0}, q_e=\fr {\mu} {d_e} + \fr {\g_e} {d_e} \fr 1 {R_0}&  \text{when  }R_0 \ge 1. \ec $$
 This checks
with the particular case in \cite{allen2012extinction}, where $\sd=1$.

\beR It is not clear intuitively why separating the transition rates into those of $F$ (which increase the norm of $x$) and those of $V$  (which do not increase the norm of $x$)
should matter for determining the extinction probabilities,  as happens in  \eqr{qeq}.  This seems an interesting question. \eeR

\sec{Rank one host-only models with pathogen, and $R_0$ readable from the flow-chart}
\subsection{The  SEIARW model with ``catalyzing pathogen" of \cite{guo2022computing} has rank one \NGM\ and   $R_0=R_J$} \label{s:Guo}

\cite{guo2022computing}  attempted to offer a ``
 {\bf definition-based method}"  for ``computing $R_0$ of dynamic models of single host species, which is
mutually coherent with the next-generation method (NGM)" (and somewhat unclear  for "computing $R_0$ for a population with multi-group models"). Unfortunately, these authors do not seem aware  of the fact that  all the single host species they examined have \NGM\ of rank one, and that in this case  there exists a simple general formula  \cite{Arino,AABBGH,AAGH}, which is  also related to the  definition-based method of \cite{Camino}.

We review now the   SEIAR model (susceptibles, exposed, infected, asymptomatic and recovered), to which  \cite{guo2022computing} add  also a pathogen  \com\ $W$, resulting in the SEIARW model. 
\be{Guo} \bc
\bc
e' = s \left(a \b_a+i \b_i +w \beta _w\right)-e d_e, &d_e=   e_i +e_a+\mu   \\
i' = e e _i-i d_i, &d_i=\g_i+\mu +\delta,\\
a'= e e_a-a d_a, &d_a=\g_a+\mu,\\
w'=a \ep_a+i \ep_i-w d_w\\
r'=a \g_a+\g_i i-\mu  r\ec\\
s'=\Lambda -s \left(a \b_a +i \b_i +w \beta _w+\mu\right).
 \ec \ee
 In matrix form, the disease equations are:
 \bea \ \bep e'\\i'\\a'\\w'\\r' \eep=
   \bep - d_e&s \b_i &s  \b_a&
  s  \beta _w&0\\
  e_i&- d_i&0&0&0\\
 e_a &0&- d_a&0&0\\
0& \ep_i& \ep_a&- d_w&0\\
0&\g_i & \g_a& 0& -\mu \eep
\bep e\\i\\a\\w\\r \eep
\eea

  In the absence  of pathogen,
  SEIAR is a rank one "generalized stage-structured infectious disease model" as revealed by its ${F=}\left(
\begin{array}{ccc}
 0 & s\beta_i  & s\beta _a \\
 0 & 0 & 0 \\
 0 & 0 & 0 \\
\end{array}
\right)
$  and by its $V=\left(
\begin{array}{ccc}
 d_e  & 0 & 0 \\
 -e_i & d_i  & 0 \\
 -e_a & 0 & d _a  \\
\end{array}
\right)$ matrix, which is triangular (compare to  \cite[Sec. 3]{AllenVdD}).

The $R_0$  has a very intuitive, and easily explainable form:
 \be{R0si} R_0= \fr \sd {d_e}\pp{ \b_i \fr {e_i}{d_i} + {\b_a} \fr {d_a} {e_a}} \ee
 (compare to  \cite[Sec. 3]{AllenVdD}, to see the general pattern for more stages).
 \beR Note that this result may be obtained with $\mI=\{e,i,a,r\}$, and also
 with $\mI=\{e,i,a\},$  which raises the question of defining the concept of minimal or "sufficient disease" set, in such a way that it allows deriving both $R_0$ and the extinction probabilities. \eeR

 After the addition of the catalyzing pathogen, the SEIARW is still  a rank one "generalized stage-structured infectious disease model", but the $R_0$ is less intuitive
\be{R0G} R_0= \fr \sd{d_e}\pp{\b_i \fr {e_i}{d_i} + {\b_a} \fr {d_a} {e_a}+
 \fr{\b_w}{ d_w}\pr{\ep_i \fr{ e_i}{d_i}+ \ep_a\fr{ e_a}{d_a}}};\ee
 {still, it may be read out of the  flow chart  ``almost by inspection"} (see  also \cite{guo2022computing} for an algorithm computing this).
\figu{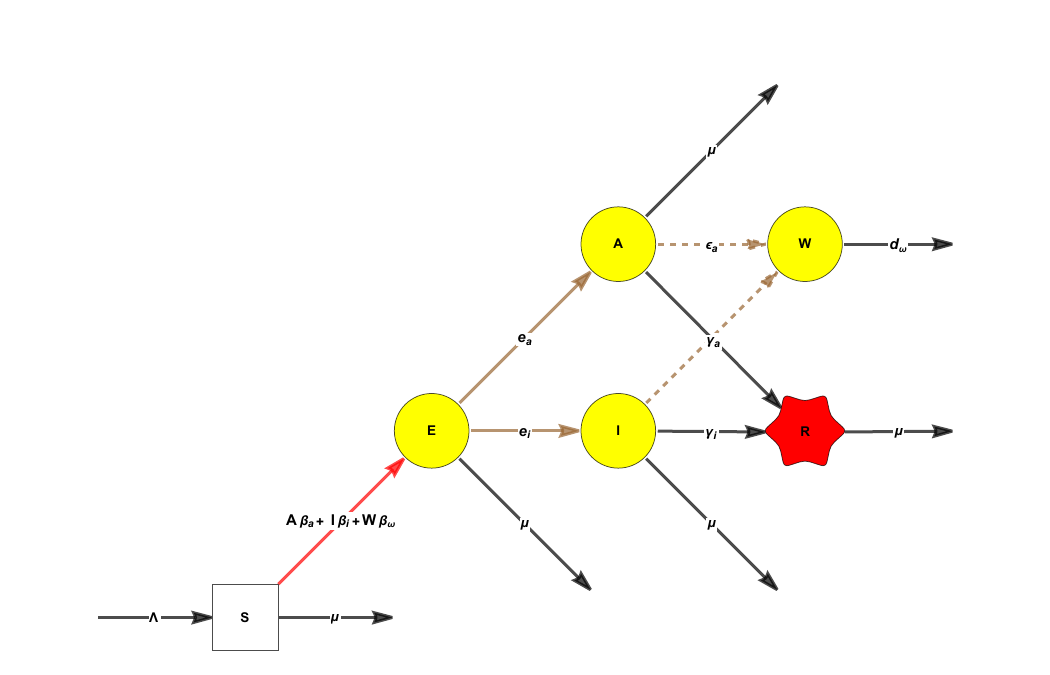}{Flow chart corresponding to the SEIARW model \eqr{Guo}.}{.8}

  Despite the fact that the \ch\ is not of Descartes type, all our three $R_0$ recipes yield the above result. We provide now details for the NGM method. After
removing the \com\ $r$ (since it does not appear in the other equations), the calls ``inf=Range[4];DFE[SEIARW,inf]; NGM[SEIARW,inf]" of our scripts yield that the DFE is
$$\left\{s\to \frac{\Lambda }{\mu },e\to 0,i\to 0,a\to 0,w\to 0\right\}$$ and
\be{FVGuo} F= s \left(
\begin{array}{cccc}
 0 & \beta_i &  \beta _a & \beta _w \\
 0 & 0 & 0 & 0 \\
 0 & 0 & 0 & 0 \\
 0 & 0 & 0 & 0 \\
\end{array}
\right)=s  \bep 1\\ 0\\0\\0\eep \bep 0 & \b_i & \b_a& \b_w\eep, V=\left(
\begin{array}{cccc}
 d_e & 0 & 0& 0 \\
 -e_i & d_i  & 0 & 0 \\
 -e_a   & 0 & d_a  & 0\\
 0 & -\ep_i  & -\ep_a  & d_w  \\
\end{array}
\right).\ee

Here the dominant eigenvalue of  $K=F V^{-1}$, that of
the transpose
$$ K^t=s\left(
\begin{array}{cccc}
 \frac{\b_i d_w e_i d_a +
 \beta _a e_a d_w d_i +\beta _w \left(e_i \ep_i
 d_a+e_a \ep_a  d_i \right)}
 {d_e d_ i d _a d_w } &
 \frac{ \b_i d_w+\ep_i \beta _w}{d_w d_i} &
  \frac{\beta _a d_w+\ep_a \beta _w}
 {d_w d_a} &
  \frac{ \beta _w}{d_w} \\
 0 & 0 & 0 & 0 \\
 0 & 0 & 0 & 0 \\
 0 & 0 & 0 & 0 \\
\end{array}
\right),$$
the rank 1 formula of \cite{Arino,AABBGH,AAGH}
$$(1,0,0,0).(V^t)^{-1}. (0,\b_i, \b_a, \b_w)^t,$$ as well as  the Jacobian factorization confirm all the result \eqr{R0G} of \cite{guo2022computing}.

\section{Target-infection-virus models}
\subsection{Two admisible $(F,V)$ decompositions and $R_0$'s for the three dimensional  model of \cite{Leen}}\label{s:Lee}
The three dimensional model of \cite[1]{Leen} is:
\be{Leen} \bc x'=\Lambda(x) -\beta x   v -\beta_{xy}  x y , \quad \Lambda(x)=\mu _x(\xd- x) 
\\\bep y'\\v'\eep =\bep \beta   x v+ \beta_{xy}  x y  \\0 \eep  -\bep \mu _y y\\
\mu _v v +\beta_{xv}   x v + \beta_{yv}   y v -b \mu _y  y \eep, \ec \ee
where we represented already the infectious equations as a difference of "new (positive) infection" terms and "transfers". The DFE is $x= \xd,y= 0,v= 0$.

This reduces  to the case with 0-delays in \cite[5.1]{YangRuan} when the rate of viruses moving into a healthy cell  $\beta_{xv}$ and the rate of viruses  moving into an infected cell  $\beta_{yv}$ are both $0$, and to the case in \cite{AAA} when $\beta_{yv}=0=\beta_{xy}$ (the latter is the cell to cell infection rate), and $\beta_{xv}=\b$.

The gradient of the infectious equations
is \be{xyvM}M= \left(
\begin{array}{cc}
 x \beta_ {xy}-\mu _y & \beta  x \\
 b \mu _y-v \beta_ {yv} & -\mu _v-x \beta_ {xv}-y \beta _{yv}
\end{array}
\right)\ee

Calling  our  NGM script
with  ``inf=\{2,3\}"   yield's \cite{Leen}'s result: namely,
$$F=x_{dfe} \left(
\begin{array}{cc}
 \beta _{xy} & \beta  \\
 0 & 0 \\
\end{array}
\right), -V=\left(
\begin{array}{cc}
 -\mu _y & 0 \\
 b \mu _y & -\mu _v-\xd \beta_ {xv}
\end{array}
\right),$$  the \NGM\ (NGM) of the infectious coordinates  at the DFE
\[K= \left(
\begin{array}{cc}
 \frac{\beta  b x}{\mu _v+x \beta _{xv}}+\frac{x \beta _{xy}}{\mu _y} & \frac{\beta  x}{\mu _v+x \beta _{xv}} \\
 0 & 0 \\
\end{array}
\right),\]
and  that the DFE  is Lyapunov-Malkin  stable when
 $R_0$ defined in
 \be{R0} R_0= \frac{x_{dfe}}{x_c}+\beta  b \frac{ x_{dfe}}{\mu _v+x_{{dfe}} \beta _{xv}}, \; \; x _c:=\fr{\mu _y}{\beta _{xy}}, \ee
is smaller than $1$, and unstable when $R_0>1$.

The Jacobian factorization provides the same  formula, despite the fact that the \ch\ is  of Descartes type only conditionally, when $\beta _{xv}\ge \beta _{xy}$.

\beR   Interestingly, another admissible decomposition $\mF=\bep \beta   x v+ \beta_{xy}  x y  \\beta \mu _y y \eep$, appears in an earlier version of \cite{Leen} at https://people.clas.ufl.edu/pilyugin/files/cosner60-dcdsB.pdf :
\be{FV} F= \left(
\begin{array}{cc}
 x \beta _{xy} & \beta  x \\
 b \mu _y & 0 \\
\end{array}
\right), \quad V=\left(
\begin{array}{cc}
 \mu _y & 0 \\
 0 & \mu _v+\beta_{xv}  \xd \\
\end{array}
\right), K=\left(
\begin{array}{cc}
 \frac{\xd }{x _c} & \frac{\beta  \xd}{\mu _v+\beta_{xv}  \xd} \\
 b & 0 \\
\end{array}
\right)
\ee

This second decomposition yields a different $R_0$:
\be{R0L}R_0'=\frac {x_{dfe}} {2 x_c}
\pr{1+
\sqrt{1+\frac{4 \beta  b x_c^2}{x_{dfe}(x_{dfe} \beta _{xv}+\mu _v)}}}.\ee

Furthermore, this early version   also shows that the two decompositions have the same stability domain for the DFE, which may be reexpressed  as
\be{stD}  R_0 =K_{1,1} + K_{1,2} K_{2,1}=\fr{ \xd}{x_c}
 +b \beta  \fr{ \xd}{\mu_v+ \beta_{xv}\xd} <1. \ee
We note that this equivalence  
 follows also by applying  the first criterion in \cite{yang2015} (when the characteristic polynomial, given here by
$\lambda^n - a_1 \lambda^{n-1} - a_2 \lambda^{n-1} -...$ has all coefficients non-negative, than $\sum_i a_i$ may be used as threshold  parameter instead of $R_0$),    with $n=3, a_1= K_{1,1}, a_2= K_{1,2} K_{2,1}, a_3=0$.
\beR
Note the second decomposition has one more non-zero term in $F$, which does not appear in ours, since we view it as a transfer and not as an interaction.  We see here an excellent example of non-uniqueness, where one must chose between an   answer  with $F$ of lower rank and a simpler $R_0$ formula,  but which is valid only under certain conditions (that the non-diagonal term $b \mu_y$ in $V$ is small enough),  and an answer with simpler $V$, which requires less assumptions on the parameters, but yields a more complicated $R_0$. \eeR

\eeR

  \beR {\bf The domain of stability, in terms of the parameters}. As an aside,  it is easy to show that \eqr{stD} is equivalent to
  \be{2in} \xd  <x_c, b < b_0:=\frac{\mu _v+\beta_{xv}  \xd }{\beta  \xd} \left(1-\fr {\xd }{x_c}\right),
\ee
where $b_0$ is the solution of the equation $R_0(b)=1$.
Thus, stability of the DFE is equivalent to both $\xd$ and the ``burst parameter" $b$ being small enough.
\eeR

 We offer now  a third gradient decomposition, which turns out to be inadmissible sometimes, but yields again our recipe's $R_0$.  Taking $ F= \left(
\begin{array}{cc}
 \xd \beta _{xy} & 0 \\
 b \mu _y & 0 \\
\end{array}
\right)$ yields $$V=F - M=\left(
\begin{array}{cc}
 \mu _y & -\beta  \xd \\
 0 & \mu _v+\xd \beta_{xv} \\
\end{array}
\right), V^{-1}=\left(
\begin{array}{cc}
 \frac{1}{\mu _y} & \frac{\beta  \xd}{ \mu _y(\mu _v+\xd \beta _{\text{xv}} )} \\
 0 & \frac{1}{\mu _v+\xd \beta _{\text{xv}}} \\
\end{array}
\right).$$
Note that $V$ is a subgenerating matrix only if $\xd \le \fr{\mu _y}{\b}$.

However   $K=\left(
\begin{array}{cc}
 \frac{\xd \beta _{xy}}{\mu _y} & \frac{\beta  \xd^2 \beta _{xy}}{\mu _v \mu _y+\xd \beta _{xv} \mu _y} \\
 b & \frac{\beta  b \xd}{\mu _v+\xd \beta _{xv}} \\
\end{array}
\right)$ yields the   correct  

$$R_0=\max \left(0,\frac{\beta  b \xd}{\mu _v+\xd \beta _{xv}}+\frac{\xd \beta _{xy}}{\mu _y}\right).$$

In the current example, the RUR algorithm works as well. The difference of the two positive terms is
$$\beta  b \xd \mu _y+\xd \beta _{ {xy}} (\mu _v +\xd \beta _{ {xv}})-\mu _y(\mu _v +\xd \beta _{ {xv}} )=\mu _y(\mu _v +\xd \beta _{ {xv}})(R_0-1)$$
for both choices $y$ and $v$ as scalar variable, and the appropriate cosmetics recovers the recipe  NGM $R_0$.

This  example illustrates the fact that sometimes several admissible and even conditionally non-admissible decompositions, as well as other approaches, may lead
to the same $R_0$.

\subsection{Two distinct approximate extinction probabilities, one for each admisible $(F,V)$ decomposition for the   model of \cite{Leen}}
The extinction probabilities of the stochastic model are of course unique. We may use 
the result of Bacaer's formula as approximations. In this interesting example, we find out that both $(F,V)$ decompositions yield reasonable results. This suggests that we have not one, but two deterministic epidemiologic approximations for a single stochastic model. This strengthens our point of view that a deterministic epidemiologic model must include a specification of the $(F,V)$ decomposition.

The respective results we got are: 
\BEN \im For the first decomposition, the extinction probabilities obtained by solving \eqr{qeq}  are
\bea
\Scale[0.8]{
\bc
q_y=1, q_z=1,  \text{when}\; R_0\leq 1,\\
q_y=\frac{\pm\sqrt{x^2 \left(\left(\beta _{\text{xy}} \left(\mu _{\text{v}}+x \left(\beta +\beta _{\text{xv}}\right)\right)+\beta  \mu _{\text{y}}\right){}^2-4 \beta  \beta _{\text{xy}} \mu _{\text{y}} \left(x \left(\beta -b \beta  +\beta _{\text{xv}}\right)+\mu _{\text{v}}\right)\right)}+x \left(\mu _{\text{v}} \beta _{\text{xy}}+\beta  \mu _{\text{y}}\right)+x^2 \left(\beta +\beta _{\text{xv}}\right) \beta _{\text{xy}}}{2 \beta  x^2 \beta _{\text{xy}}}, \text{when}\; R_0> 1\\
q_z=\frac{ \left(\mu _{\text{v}}+x \beta _{\text{xv}}\right) \left(\pm\sqrt{x^2 \left(\left(\beta _{\text{xy}} \left(\mu _{\text{v}}+x \left(\beta +\beta _{\text{xv}}\right)\right)+\beta  \mu _{\text{y}}\right){}^2-4 \beta  \beta _{\text{xy}} \mu _{\text{y}} \left(x \left(\beta -b \beta  +\beta _{\text{xv}}\right)+\mu _{\text{v}}\right)\right)}-x \mu _{\text{v}} \beta _{\text{xy}}+x^2 \left(\beta +\beta _{\text{xv}}\right) \left(-\beta _{\text{xy}}\right)+\beta  x \mu _{\text{y}}\right)}{2 \beta ^2 b x^2 \mu _{\text{y}}}. 
\ec}
\eea

\im For the second decomposition, the extinction probabilities obtained by solving \eqr{qeq}  are:
\bea
\Scale[0.8]{
\bc
q_y=1, q_z=1,  \text{when}\; R_0\leq 1,\\
q_y=\frac{\pm\sqrt{x^2 \left(\left(\beta  (b+1) \mu _{\text{y}}+\beta _{\text{xy}} \left(\mu _{\text{v}}+x \left(\beta +\beta _{\text{xv}}\right)\right)\right){}^2-4 \beta  \beta _{\text{xy}} \mu _{\text{y}} \left(\mu _{\text{v}}+x \left(\beta +\beta _{\text{xv}}\right)\right)\right)}+\beta  (b+1) x \mu _{\text{y}}+x \mu _{\text{v}} \beta _{\text{xy}}+x^2 \left(\beta +\beta _{\text{xv}}\right) \beta _{\text{xy}}}{2 \beta  x^2 \beta _{\text{xy}}},  \text{when}\; R_0> 1.\\
q_z=\frac{\left(\mu _{\text{v}}+x \beta _{\text{xv}}\right) \left(\pm \sqrt{x^2 \left(\left(\beta  (b+1) \mu _{\text{y}}+\beta _{\text{xy}} \left(\mu _{\text{v}}+x \left(\beta +\beta _{\text{xv}}\right)\right)\right){}^2-4 \beta  \beta _{\text{xy}} \mu _{\text{y}} \left(\mu _{\text{v}}+x \left(\beta +\beta _{\text{xv}}\right)\right)\right)}+\beta  (b+1) x \mu _{\text{y}}-x \mu _{\text{v}} \beta _{\text{xy}}+x^2 \left(\beta +\beta _{\text{xv}}\right) \left(-\beta _{\text{xy}}\right)\right)}{2 \beta  b x \mu _{\text{y}} \left(\mu _{\text{v}}+x \left(\beta +\beta _{\text{xv}}\right)\right)}.
\ec}
\eea
\EEN

In a numeric instance, we found the two results reasonably close to each other.

\section{Multi-strain host only   models}\label{s:mul}
Multi-strain diseases are diseases that consist of several strains, or serotypes.
One interesting thing about  multi-strain    models is that besides the DFE we have new boundary points which are relevant epidemiologically,
in which one subset of strains $A$ is present (``resident"). We have then a natural coexistence  of several ``$\mR$ thresholds":
\BEN  \im $  R_A$ is the bifurcation threshold at which the DFE stops being  stable, when the only \com s present are those of $A$.
 \im $  {\mR_{A}}$ is the bifurcation threshold at which the boundary point  $E_A$ starts existing (in the presence of the $A^c$ \com s).
\im $  R_{A^c,A}$ is the bifurcation threshold at which the boundary point $ E_A $ stops being  stable, i.e. when the $A^c$ \com s invade the  $A$ \com s.
\EEN

Note that for two strains already, we have at least two new thresholds, $R_{21}, R_{12},$ which together  with $R_0$ and the  thresholds $R_1$, $R_2$ of the individual strains  divide the line into $6$ regions with  different stability properties.
Studying the relations between the various thresholds in parameter space is quite a challenging topic.  However, their calculation is a priori of the same level of difficulty as  for the DFE.

\subsection{The  two-strain SIS tuberculosis model of \cite[Sec4.4]{Van}}
The model presented here is a limiting case of that presented in the next section, obtained when the transition rates $\g_1, \g_2$ converge to $\I$. It also generalizes
the  two-strain SIS tuberculosis model of \cite[Sec4.4]{Van} by allowing cross-infections in both directions

\bea
\bc
i_1'=i_1 \left(i_2 \left(\nu _2-\nu _1\right)+\beta _1 s-\sigma _1-b\right)=i_1 \left(i_2 \left(\nu _2-\nu _1\right)+\beta _1 s-d _1\right),\\
i_2'=i_2 \left(i_1 \left(\nu _1-\nu _2\right)+\beta _2 s-\sigma _2-b\right)=i_2 \left(i_1 \left(\nu _1-\nu _2\right)+\beta _2 s-d _2\right),\\
s'={b}-s \left(\beta _1 i_1+\beta _2 i_2+b\right)+i_1 \sigma _1+i_2 \sigma _2,
\ec
\eea
where we put $d_1=\sigma _1-b, d_2=\sigma _2-b$ in the first two equations, to simplify their notation (the last equation may be removed, since $s=1-i_1-i_2$).

Noting that the first two equations  factor yields the following three boundary steady states, where $\mathbf{x} = ( i_1,  i_2, s)$:
\begin{eqnarray*}
\mathbf{x}_0 & = & \left( 0,  0,1\right) \\
\mathbf{x}_1 & = & \displaystyle{\left(1 - \mR_1^{-1},0,\mR_1^{-1}\right)} \\
\mathbf{x}_2 & = & \displaystyle{\left( 0,
1 - \mR_2^{-1},\mR_2^{-1}\right)},
\end{eqnarray*}
where we put $$\mR_1 = \displaystyle{\frac{\beta_1}{{b+\sigma_1}}},
\mR_2 = \displaystyle{\frac{\beta_2}{{b+\sigma_2}}}.$$

The \emph{disease free steady state} $\mathbf{x}_0$ exists for all parameter values, while the \emph{original strain only steady state} $\mathbf{x}_1$ is physically relevant if and only if $\mR_1 > 1$ and the \emph{emerging strain only steady state} $\mathbf{x}_2$ is physically relevant if and only if $\mR_2 > 1$.

There may also be a fourth non-negative coexistence equilibrium (COE), given by
\be{4V}\begin{cases}
i_1= \frac{\beta _1 d_2-\beta _2 d_1-\left(\nu _1-\nu _2\right) \left(\beta _2-d_2\right)}{\left(\nu _1-\nu _2\right) \left(\beta _1-\beta _2+\nu _1-\nu _2\right)}\\
i_2=\frac{d_1 \left(\beta _2-\nu _1+\nu _2\right)-\beta _1 \left(d_2-\nu _1+\nu _2\right)}{\left(\nu _1-\nu _2\right) \left(\beta _1-\beta _2+\nu _1-\nu _2\right)}\\
s=1-i _1-i _2.\ec
\ee

Note this depends only on $\nu _1-\nu _2,$ which shows that the case $\nu _1=0$ considered  in \cite[Sec4.4]{Van} is not that restrictive\fn[4]{However, the appearance
of $\nu _1-\nu _2$  in the denominator suggests limiting diffusion phenomena, which may be worth studying in their own right.}. In this case, the COE point simplifies to:
\be{4R}\begin{cases}
i_1= \frac{d_2 \left(d_1 \left(\mathcal{R}_1-\mathcal{R}_2\right)+\nu  \left(\mathcal{R}_2-1\right)\right)}{\nu  \left(-d_1 \mathcal{R}_1+d_2 \mathcal{R}_2+\nu \right)}\\
i_2=\frac{d_1 \left(d_2 \left(\mathcal{R}_2-\mathcal{R}_1\right)+\nu(1 -\mathcal{R}_1)\right)}{\nu  \left(-d_1 \mathcal{R}_1+d_2 \mathcal{R}_2+\nu \right)}\\
s=1-i _1-i _2\ec,
\ee
which is positive iff {$\mR_2>1$ and the following conditions hold}
\be{nuR}
 \bc
 \mR_1>\frac{\nu+\mR_2 d_2}{\nu+d_2},\; 0<\nu<\frac{d_1(\mR_2-\mR_1)}{\mR_2-1},  \; \mbox{or},\\
\mR_1<\frac{\nu+\mR_2 d_2}{\nu+d_2}, \pr{0<d_1<\nu(1-\frac{1}{\mR_2}) \; \mbox{or}\; d_1>\nu(1-\frac{1}{\mR_2}), \nu<\frac{d_1(\mR_2-\mR_1)}{1-\mR_2}}.
\ec
\ee

We give now some details of the NGM implementation for the  three boundary points. Recall  that the idea  is to project the ODE at each boundary point on the $0$ coordinates (or some subset), while fixing the other coordinates.   We must  compute therefore new $(F,V)$ pairs at each boundary point, since the respective zero coordinates are different.

\BEN
\im
At the DFE, the zero coordinates are   $\left\{i_1,i_2\right\}$, and so $\mI=\{1,2\}$.

Our script yields
the expected result

$$R_0=Max\pp{\frac{\beta _2 s_{dfe}}{\sigma _2+b},
\frac{\beta _1 s_{dfe}}{\sigma _1+b}}=Max\pp{R_1,R_2}, R_i=\sd \mR_i= \mR_i, i=1,2.$$

\im
At  $\mathbf{x}_2$,
$\mI=\{1\}$, and
$$R_{12}= \frac{\mR_1}{\mR_2}+\frac{\left(\nu _2-\nu _1\right) \left(1 -\mR_2^{-1}\right)}{b+\sigma _1}.$$

When $\nu_1=0, \nu_2=\nu,$  we recover the  result  \cite[(18)]{Van}
$R_{12}= \frac{\mR _1  }{\mR _2  } + \frac{\nu  }{b+\sigma _1} \left(1 -\mR_2^{-1}\right).$

This implies that stability holds iff $\mR_2 >1$ and $\mR_1$ is not too big, more precisely: \be{Vs} R_{12}<1 \Eq
\mR_1< \mR_2+ \frac{\nu  \left(1-\mR_2\right)}{b+\sigma _1}.\ee

For  a sanity check, we will derive the stability condition of the point $\mathbf{x}_2$ also by the direct Jacobian approach.  The Jacobian
 at $\mathbf{x}_2$ is
$$\left(
\begin{array}{ccc}
 \frac{-\beta _2 \left(b+\nu _1-\nu _2+\sigma _1\right)+\beta _1 \left(b+\sigma _2\right)+\left(\nu _1-\nu _2\right) \left(b+\sigma _2\right)}{\beta _2} & 0 & 0 \\
 -\frac{\left(\nu _1-\nu _2\right) \left(b-\beta _2+\sigma _2\right)}{\beta _2} & 0 & -b+\beta _2-\sigma _2 \\
 \sigma _1-\frac{\beta _1 \left(b+\sigma _2\right)}{\beta _2} & -b & \sigma _2-\beta _2 \\
\end{array}
\right).$$
In the case of \cite{Van},
the eigenvalues are
$$\left\{-b,-\left(\left (\mR_2-1\right) \left(b+\sigma _2\right)\right),\frac{(b + \sigma _1) \left (\mR_1-\mR_2 \right)+\nu  \left(\mR_2-1\right)}{\mR_1}\right\}.$$

The second eigenvalue is negative iff $\mR_2>1$,
and the third eigenvalue is negative when
$$ \left (\mR_1-\mR_2 \right)+\fr \nu  {b + \sigma _1}\left(\mR_2-1\right)<0 \Eq R_{12} <1\text{ see } \eqr{Vs}.$$

\im

An analog result holds by symmetry at  $\mathbf{x}_1$, where $\mI=\{2\}$, and
{$$ \mR_{21}= \frac{\left(\nu _1-\nu _2\right) \left(\mathcal{R}_1-1\right)}{d_2 \mathcal{R}_1}+\frac{\mathcal{R}_2}{\mathcal{R}_1}.$$}
\EEN

We illustrate now via a $i_1$ bifurcation diagram that,  as natural,   when $\beta_1$ is  small enough,  the $x_2$ fixed point is stable, to be replaced  as attractor first by the COE, and finally by the $x_1$ fixed point, when $\beta_1$ increases.

\figu{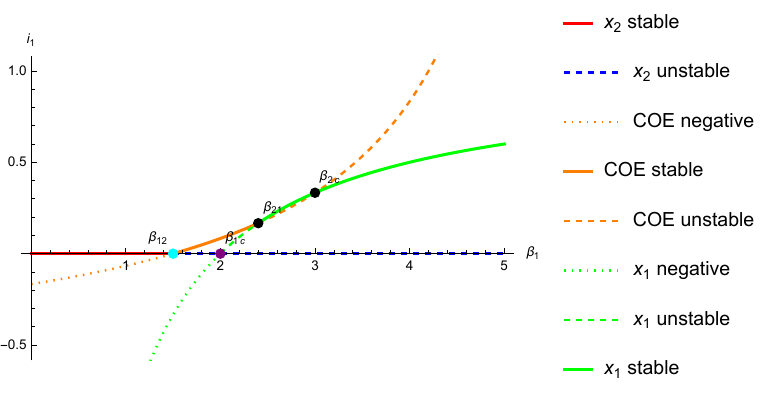}{$i_1$ bifurcation diagram   when $\beta_1$ varies and $\nu_1=0, \nu_2=\nu=3=\beta_2=3, b=\sigma_1=\sigma_2=1,  R_1=\fr {\b_1} 2, R_2=\fr 3 2,$ so that $x_2$
 is always positive. Since $ R_0 \ge R_2 >1$, the DFE is never stable. Observe the following three regimes: a) until  $\beta_{12}=1.5$ defined by equality in  ${R_{12}:= \frac{\nu  \left(\mathcal{R}_2-1\right)}{d_1 \mathcal{R}_2}+\frac{\mathcal{R}_1}{\mathcal{R}_2}}\le 1 \Eq \beta_{12}\le \frac{\beta_2 \left(b-\nu +\sigma _1\right)}{b+\sigma _2}+\nu,$ the only stable solution is $x_2$.
b) At $\beta_{12}=1.5$, $x_2$ becomes unstable
and the coexistence  solution becomes nonnegative and stable, until $\beta_{21}$ defined by $\mR_{21}=\frac{\mathcal{R}_2}{\mathcal{R}_1}-\frac{\nu  \left(\mathcal{R}_1-1\right)}{d_2 \mathcal{R}_1}=1 \Eq \beta_{21}=\frac{\left(b+\sigma _1\right) (\beta_2+\nu )}{b+\nu +\sigma _2}=2.4$. This is also  the first intersection point of the COE and $x_1$.  For a numerical check, at  $\beta_{1c}=2$,  defined by $\mR_1=1 \Eq \beta_{1c}=b+\sigma_1,$ where the $x_1$
 solution emerges  and is initially {unstable}, the  eigenvalues for the COE are $\pr{-1, -0.333333 \pm 0.235702 Im}$.
c) After $\b_1=\beta_{21} \Eq \mR_{21} <1$,   the $x_1$ solution becomes stable and  the COE loses its stability (the latter was checked numerically). Note that at $\beta_{2c}=3 \Eq \mR_1=R_{12} \Eq \beta_1=\nu$ there is no stability change: the COE and $x_1$ continue to be unstable 
  and stable, respectively.}{1} 

 \begin{figure}[H]
    \centering
    \begin{subfigure}[a]{0.4\textwidth}
    \centering
        \includegraphics[width=\textwidth]{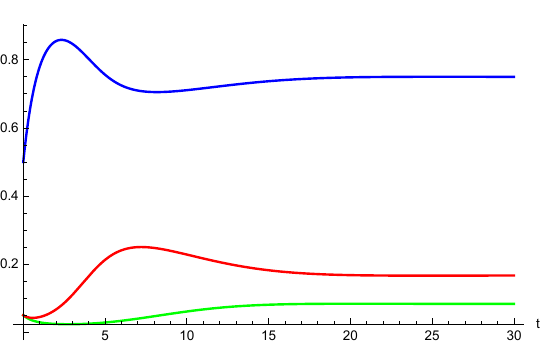}
        \caption{ $(i_1,i_2,s)$-time plot at the point $\beta_{1c}=2$ reveals convergence towards the COE }
    \end{subfigure}%
    ~
     \begin{subfigure}[a]{0.38\textwidth}
     \centering
        \includegraphics[width=\textwidth]{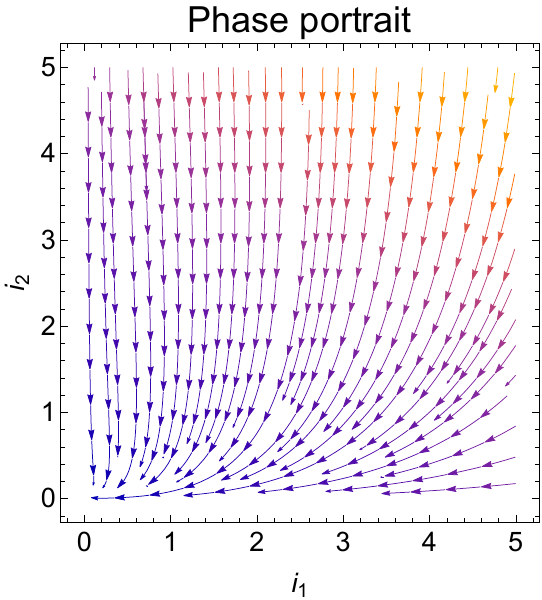}
        \caption{$(i_1,i_2)-$stream plot  }
    \end{subfigure}
    \caption{Time and phase-plot   at the point $\beta_{1c}=2$ illustrating convergence towards COE=$\pr{i_1\to 0.0833333,i_2\to 0.166667,s\to 0.75}$ .}
    \end{figure}

\subsection{The  minimal disease set of the  multi-strain host only   dengue model with antibody-dependent enhancement (ADE)  \cite{Agu08}}\label{s:inv}
The ADE (antibody-dependent enhancement) effect, believed to occur  for Dengue and  Zika, means that infection with a single serotype is asymptomatic, but
infection with a second serotype may lead to serious illness, accompanied by greater infectivity. It was first studied  mathematically by \cite{ferguson1999effect,schwartz2005chaotic}, who showed that for
sufficiently small ADE, the numbers of infectives of each serotype synchronize, with outbreaks occurring in
phase, but  when the ADE increases past a threshold, the system becomes chaotic, and infectives of each serotype
desynchronize (however, certain groupings of the primary and secondary infectives remain synchronized even
in the chaotic regime).  Subsequently, \cite{billings2008vaccinations} examined the effects of single-strain vaccine campaigns on the dynamics of an epidemic multi-strain Dengue model. We  cite now the eloquent Dengue description given by these authors:

"What makes modeling the dengue virus so interesting is that it has developed a sophisticated
spreading process. Dengue is known to exhibit as many as four coexisting serotypes (strains) in
a region. Once a person is infected and recovered from one serotype, they confer life-long immunity from that serotype. However, the antibodies that the body develops for the first serotype will
not counteract a second infection by a different serotype. In fact, due to the nature of the disease,
the antibodies developed from the first infection form complexes with the second serotype so that
the virus can enter more cells, increasing viral production.  The increased transmission rate in subsequent infections is known
as antibody-dependent enhancement (ADE).
 ADE is an alarming evolutionary development in multistrain viruses with respect to vaccines.
An optimal vaccination would need to cover all strains of the disease at once, or the vaccinations
could increase transmission of the strains not covered. This is particularly dangerous for people
who have dengue because the infections are more severe in individuals who already have dengue
antibodies."

A  multi-strain model which adds  further \com s allowing for temporary cross-immunity has been developed in the works of Aguiar, Stollenwerk and Kooi \cite{aguiar2007new,Agu08,aguiar2009torus,
stollenwerk2017hopf,aguiar2022mathematical}.

In this section we consider a  ten \com s asymmetric version of the model of  \cite{Agu08}, whose variables, denoted  by capital letters, represent:
 \BEN \im  $S$ are  individuals susceptible to both strains;
 \im $I_i$, for $i,j=1,2$ are individuals infected  with strain $i$ and with temporary cross-immunity to strain $j\neq i$;

 \im $R_i$    are  individuals who have
recovered from strain $i$, but are not yet  susceptible to the other strain $j$;
\im $S_i$  are individuals who have
recovered from strain $i$, and have become susceptible to the other strain $j$;
\im $Y_{j}=I_{ij}$  are  individuals previously infected with strain $i$  and now immune to it,  but  reinfected with strain $j, i,j=1,2, i\neq j$;
\im  $R$, omitted in \eqr{Agu} since they do not feed back to the other components, are the recovered individuals immune  to all the strains.

\EEN

After  denoting by small letters the corresponding proportions, we arrive at:
\be{Agu}
\bc s'=\mu -s \left(\beta _1 i_1+\beta _2 i_2+\mu +\beta _1 y_1 \phi _1+\beta _2 y_2 \phi _2\right),\\
i_1'=\beta _1 s \left(i_1+y_1 \phi _1\right)-i_1 \left(\gamma _1+\mu \right),\\
r_1'=\gamma _1 i_1-r_1 \left(\theta _1+\mu \right),\\
s_1'=\theta _1 r_1-s_1 \left(\beta _2 \alpha  _2 \left(i_2+y_2 \phi _2\right)+\mu \right),\\
{y_2'}=\beta _2 \alpha  _2 s_1 \left(i_2+y_2 \phi _2\right)-y_2 \left(\gamma _2+\mu \right),\\
{i_2'}=\beta _2 s \left(i_2+y_2 \phi _2\right)-i_2 \left(\gamma _2+\mu \right),\\
{r_2'}=\gamma _2 i_2-r_2 \left(\theta _2+\mu \right),\\
s_2'=\theta _2 r_2-s_2 \left(\beta _1 \alpha  _1 \left(i_1+y_1 \phi _1\right)+\mu \right),\\
y_1'=\beta _1 \alpha  _1 s_2 \left(i_1+y_1 \phi _1\right)-y_1 \left(\gamma _1+\mu \right).\ec
\ee

Besides the DFE where $s=1$ and all the other \com s are $0$, this system has also two other boundary points. With
$\mR_i=\frac{\beta _i}{\gamma _i+\mu }$, these are: \BEN \im one with $i_2=r_2=s_2=y_1=y_2=0$, given by
\bea E_1=\pr{\fr \mu{\beta _1}  \left(\mR _1-1\right),
\frac{\mu \gamma _1  }{\beta _1 \left(\alpha _1+\mu \right)} (\mR _1-1),
\frac{ \alpha _1 \gamma _1}{\beta _1
\left(\alpha _1+\mu \right)}(\mR _1-1),0, 0,0,0,0,\frac{1}{\mR _1}}
\eea
\im and one with $i_1=r_1=s_1=y_1=y_2=0$, given by
\bea E_2=\pr{0,0,0,0,\fr \mu{\beta _2}  \left(\mR _2-1\right),
\frac{\mu \gamma _2  }{\beta _2 \left(\alpha _2+\mu \right)} (\mR _2-1),  \frac{ \alpha _2 \gamma _2}{\beta _2 \left(\alpha _2+\mu \right)}(\mR _2-1),0,\frac{1}{\mR _2}}\\
\eea
\EEN

Thus, $\mR_i, i=1,2$ are the bifurcation values at which these two boundary points appear.

The maximal disease set contains  $I_i,R_i,S_i,Y_i, i=1,2$. The DFE may be determined already using the  disease set $I_i,Y_i, i=1,2$, which has the advantage of possessing a simple
 \ch\ with two factors $R_1(X),R_2(X)$, which yields:
$$R_J=\max [R_1(X),R_2(X)], R_1(X)=\frac{\beta _2 \left(\alpha _2 s_1 \phi _2+s\right)}{\gamma _2+\mu }, R_2(X)=\frac{\beta _1 \left(\alpha _1 s_2 \phi _1+s\right)}{\gamma _1+\mu }.$$

Also,
 our scripts find that
\be{R0I} R_{ji}=\sd \; \mR_j, j\neq i, i=1,2.\ee
Finally, applying the NGM script to $E_i, i=1,2$ yields the elegant relation
\be{R0A} R_0=\sd \max\pp{ \mR_1, \mR_2}=\max\pp{ R_{21}, R_{12}}.\ee

\beR
Note the notations $R_1(X),R_2(X)$, suggesting that we want to view these as  polynomials
in the variables of the model, rather than  values evaluated at one of the fixed
points.
\eeR

We end this section by drawing the attention to the object which allowed computing  the key polynomials
  $R_1(X), R_2(X)$.
  \beD A) A minimal disease set $\mI$ is a minimal set which still allows the computation
  of the DFE, after being set to $0$.

  B) The model factors are the factors which may admit positive roots in the \ch\ of the Jacobian with all variables in $\mI$ set to $0$.

  \eeD

\beR
Assume  \wl $\mR_1 < \mR_2$. Two situations may arise:
\bea \bc \sd \mR_1 < \mR_1 < \sd \mR_2 < \mR_2\\
 \sd \mR_1 < \sd \mR_2 < \mR_1  < \mR_2,\ec \eea
 and in each of them $1$ may lie in any of the  partition  intervals. This gives raise to $6$ disjoint cases:
 \be{cl}\bc \mR_1 <  \mR_2 \le 1& \text{the DFE is the only boundary equilibrium}\\ \sd \mR_1 < \sd \mR_2 < 1<\mR_1  < \mR_2&\text{both $E_1,E_2$ exist and are unstable}\\
 \sd \mR_1 <1< \min[\mR_1 , \sd \mR_2] < \mR_2&\text{$E_1$ unstable, $E_2$ stable}\\
 \sd \mR_1 < \mR_1 < 1<\sd \mR_2 < \mR_2&\text{only $E_2$ exists and is stable}\\
 \sd \max[\mR_1 , \sd \mR_2] <1< \mR_2&\text{only $E_2$ exists and is unstable}
 \\1< \sd \mR_1 <  \sd \mR_2& \text{competition between the two stable dominant} \\
    & \text{strains } E_1,E_2.\ec \ee

  All these cases  have been investigated in detail, for  a more general model, in \cite{Bulh}, reviewed in the next section;   it turns out that the results are fully determined by the model factors.\eeR

  Before proceeding, let us give a name to the very interesting structure we have started to investigate.

\beD \label{d:sms} A Descartes multi strain model of order $M$ is an epidemic model for which
the characteristic polynomial of the Jacobian factors completely over the rationals as a product of terms, precisely $M$ of which are ``Descartes polynomials".
For such models, the Jacobian factorization threshold is defined as
$$R_J(X):=\max_{1\le m \le M} R_m(X). $$
\eeD

One may check that
\beL \label{l:De} For Descartes multi strain models of order $K$,  the local stability set is a subset of
$$R_J(X) \leq 1.$$
\eeL

\beR The example of this section is a Descartes two -strain model (since the \ch\ has only linear factors, precisely $2$ of which have constant coefficient which may change sign).
\eeR

\subsection{Effects of single-strain vaccination on the dynamics of a    multi-strain host only dengue model with ADE}\label{s:Bulh}

In this section we will show that the mysterious formula \eqr{R0A} continues to
hold under the considerably more complicated two strains model of \cite{Bulh}, with vaccination applied to one strain only.
The model studied in  \cite{Bulh} is depicted in Figure \ref{fig:diagrama}.
\begin{figure}[hbt!]
\centering
\includegraphics[scale=0.6]{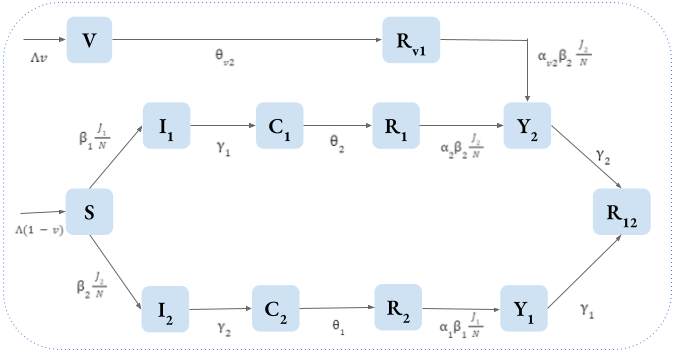}
\caption{Schematic representation of the infection status due to the concomitant transmission of viruses $1$ and $2$, considering that the population is vaccinated against the virus $1$.}
\label{fig:diagrama}
\end{figure}

This model involves twelve \com s, two of which capture the vaccination against  strain $1$.
 \BEN \im  $S=S_0$ are  individuals susceptible to both strains;
 \im $I_i$, for $i,j=1,2$ are individuals infected  with strain $i$ and with temporary cross-immunity to strain $j\neq i$;
 \im $C_i$ ($R_i$ in the original model of \cite{aguiar2007new}) are individuals
recovered from strain $i$ and hence permanently
immune  to  it, and with temporary cross-immunity to strain $j\neq i$;
 \im $R_i$  ($S_i$ in the original model of \cite{aguiar2007new})  are unvaccinated individuals who have
recovered from strain $i$, but have become susceptible to the other strain $j$;
\im $Y_{j}$ ($I_{ij}$ in the original model of \cite{aguiar2007new}) are  individuals previously infected with strain $i$  and immune to it,  but  reinfected with strain $j, i,j=1,2, i\neq j$;
\im  $R=R_{12}$ are individuals immune  to all the strains;
\im Finally, there are individuals $V$ who are vaccinated    against strain $1$ and still susceptible to strain $2$, and individuals   $R_{v1}=S_{v}=Z$ who have been vaccinated against strain $1$ and subsequently became infected by strain $2$.
\EEN
 Denote by $N(t)= S(t) + V(t) + I_1(t) + I_2(t) + C_1(t) + C_2(t) + R_1(t) + R_2(t) + Y_1(t)+Y_2(t)+S_{v}(t) + R_{12}(t)$ the total population, put   $J_i=I_i+ Y_i, \quad i=1,2$, and assume  that the two forces of infection acting on $S$ are:
\begin{equation}\nonumber
F_i=\beta_i \fr{J_i}N,
\end{equation}
and that the  forces of infection acting on $Y_i=S_i, i=1,2$
are: \begin{equation}\nonumber
\al_1 \beta_1 \fr{J_1}N, \al_2 \beta_2 \fr{J_2}N, \al_{v} \beta_2 \fr{J_2}N,
\end{equation}
where $\al_1,\al_{2},\al_{v}$ denote   decreases  or increases of the susceptibility to secondary
infections ($\al_i>1$ implying an ADE effect).

The following equations, with appropriate initial conditions, represent the disease dynamics model:

\begin{align}
\dfrac{dS}{dt}&= (1-\xi)\mu - \beta_1J_1\frac{S}{N}-\beta_2J_2\frac{S}{N} -\mu S \nonumber\\
\dfrac{dI_1}{dt}&=\beta_1J_1\frac{S}{N}-(\gamma_1+\mu)I_1\nonumber  \\
\dfrac{dC_1}{dt}&=\gamma_1I_1-(\theta_1+\mu)C_1 \nonumber \\
\dfrac{dR_1}{dt}&=\theta_{2}C_1 -\al_2 \beta_2J_2\frac{R_1}{N}-\mu R_1 \nonumber \\
\dfrac{dY_2}{dt}&=\al_2\beta_2J_2\frac{R_1}{N}+\al_{v}\beta_2J_2\frac{S_{v}}{N}-(\gamma_2+\mu)Y_2 \nonumber \\
\dfrac{dI_2}{dt}&=\beta_2J_2\frac{S}{N}-(\gamma_2+\mu)I_2 \nonumber \\
\dfrac{dC_2}{dt}&=\gamma_2I_2 -(\theta_2+\mu)C_2  \nonumber \\
\dfrac{dR_2}{dt}&=\theta_{1}C_2 -\al_1 \beta_1J_1\frac{R_2}{N}-\mu R_2 \nonumber\\
\dfrac{dY_1}{dt}&=\al_1\beta_1J_1\frac{R_2}{N}-(\gamma_1+\mu)Y_1
\nonumber \\
\dfrac{dV}{dt}&= \xi \mu - (\theta_{v}+\mu)V \nonumber\\
\dfrac{dS_{v}}{dt}&=\theta_{v}V-\al_{v} \beta_2J_2\frac{S_{v}}{N}-\mu S_{v} \nonumber
\\
\dfrac{dR_{12}}{dt}&= \gamma_1Y_1+\gamma_2Y_2-\mu R_{12}\label{eq:equations}
\end{align}

Table \ref{tab:parameters} summarizes the parameters and compartments of the model.

\begin{table}[H]
\centering
\caption{Parameters and compartments of the model.}
\scalebox{0.78}{
\begin{tabular}{|c|c|}
\hline
\textbf{Parameter} & \textbf{Description (for $i, j = 1, 2$)}\\
\hline
$\mu$ & Birth rate \\
$\mu$ & Per capita death rate \\
$\beta_i$ & Transmission rate of virus $i$\\
$\gamma_i$ & Per capita recovery rate of infected people with virus $i$\\
$\theta_{i}$ & Per capita loss rate of cross-immunity to virus $i$ after previous infection with virus $j$ \\
$\theta_{v}$ & Per capita loss rate of cross-immunity to virus $2$ obtained by vaccination\\
$\al_i$ & ADE factor that can alter the susceptibility of unvaccinated individuals to the virus $i$ \\
$\al_{v}$ & ADE factor that can alter the susceptibility of vaccinated individuals to virus $2$\\
$\xi$ & Per capita vaccination rate\\
\hline
\textbf{Compartments} & \textbf{Description}\\
\hline
$S$ & Susceptible individuals to both virus\\
$V$ & Vaccinated individuals against the virus $1$\\
$I_i$ & Individuals with primary infection by the virus $i$ \\
$C_i$ & Individuals recovered from infection with virus $i$ and have cross-immunity to virus $j$ \\
$R_i$ & Unvaccinated individuals immune to virus $i$ and susceptible to virus $j$ \\
$Z=S_{v}$ &  Individuals vaccinated for virus $1$ and susceptible to virus $2$ \\
$Y_1$ & Individuals infected by virus $1$ and recovered and hence immune to virus $2$\\
$Y_2$ & Individuals infected by virus $2$ and immune to virus $1$ either due to recovery or vaccination \\
$R_{12}$ & Individuals immune to both virus\\
\hline
\end{tabular}}
\label{tab:parameters}
\end{table}

This system does not have negative cross effects; therefore, it leaves the non-negative quadrant invariant \cite{hun}.
It follows from the equations that $$\frac{dN(t)}{dt}=\mu(1- N(t)).$$ Therefore, \[\lim_{t\rightarrow +\infty}N(t)=1.\] Assuming  $N(0)=1$ implies that $N(t)=1$, for $t\geq 0$.
Using this, we may assume \wl that $N=1$, and work with the proportions, to be denoted by the corresponding lowercase letters.

 The only non-zero \com s in the DFE, to be denoted by $E_0$,  are easily found to be $$s_0=1-  \xi,z_0=\xi \frac{ \theta _v}{\mu +\theta _v},v_0=\xi \frac{ \mu}{\mu +\theta _v};$$
  in fact, the last value holds at any fixed point. As known from \cite{Bulh}, there are also two endemic points on the boundary, whose rather complicated formulas
  will be given later.

  \beR From a modeling point of view, this system has crucial parameters like $\al_v$ (note that $\al_v=0$ means perfect vaccination, and $\al_v=1$ means that infection by second strain  is equally likely for   vaccinated people).
  \eeR

Due to conservation, the system evolves in a compact domain, and so we may eliminate one \com, for example $V$, from the analysis. Finally, the last compartment does not send input to the others, and  therefore may also be  disregarded in the analysis.

\sssec{The  Jacobian   $R_J(X)$ is the max of two polynomials, obtained using a minimal disease set}

We may tackle this example via  the Jacobian factorization  approach, choosing
the {\bf minimal disease set} $\mI=(i_1,i_2,y_1,y_2)$, just like in the previous section.
 Again, the characteristic polynomial of the Jacobian with the variables in $\mI$ set to $0$ factors completely as a product of linear terms
 $$(\mu +u)^5 \left(\gamma _1+\mu +u\right) \left(\gamma _2+\mu +u\right) \left(\theta _1+\mu +u\right) \left(\theta _2+\mu +u\right) \left(\mu +u+\theta _v\right)\times$$
\bea  \left(\gamma _1+\mu -\al _1 \beta _1 r_2-\beta _1 s+u\right) \left(\gamma _2+\mu -\al _2 \beta _2 r_1-\beta _2 s-\beta _2 z  \al _v+u\right),
\eea
only two of which (the 7'th and 8'th factors) may yield positive eigenvalues.   Both are of \Dt, and instability may occur iff
\be{Jmax}
R_J(X):=\max[R_1(X),R_2(X)]=\max[ \frac{\beta _1 \left(\al _1 r_2+s\right)}{\gamma _1+\mu },
\frac{\beta _2 \left(\al _2 r_1+z  \al _v+s\right)}{\gamma _2+\mu }]>1.\ee

At the DFE, $r_1=r_2=0,$ and this yields
\be{R0B} R_J:=R_J(E_0)=R_N=\max[ s_0  \fr{\beta _1}{d_{1}},
s_0  \fr{\beta _2 }{d_{2}}+ z_0  \fr{\beta _2 \al _v}{d_{2}}], \quad d_{1}=\gamma _1+\mu, d_{2}=\gamma _2+\mu.\ee
This expression reveals a pattern similar to \eqr{R0G}, with the difference that the existence of two strains are reflected in the max,
and that the second strain is alimented by two classes of susceptibles, one of which is the people vaccinated against the first strain.

In addition to the disease-free equilibrium, there might exist  two more equilibriums on the boundary: the endemic equilibrium where there are only infections by the strain $1$, $E_1$, and the endemic equilibrium where there are only infections by the strain $2$, $E_2$, reviewed in the next section.



\subsubsection{The endemic boundary equilibrium $E_i$ exist iff $R_i(E_0)>1$}

At the equilibrium $E_1$, the values of $I_2,C_2,R_2,Y_1,Y_2$ and $R_{12}$ are zero.
The coordinates are easily found by the ``Solve"  command. Those of $V,Z$ are the same as at the DFE,  and the others are:
\begin{eqnarray} \label{eq:E1}s_1 =\frac{\gamma _1+\mu }{\beta _1},  \; i_1 =\frac{\mu }{\beta _1}\pp{ \fr{1-\xi}{s_1 }-1}:=\frac{\mu}{\b_1}\left({\mathcal{R}_1}-1\right), \;
    c_1 =\frac{\gamma_1}{\theta_1+\mu}i_1 , \; r_1 =\frac{\theta_1}{\mu}c_1
\end{eqnarray}
{ where }  \begin{equation}\label{eq:R1}\mathcal{R}_1=(1-\xi)\frac{\beta_1}{\gamma_1+\mu} =R_1(E_0)\end{equation}
(the endemic equilibrium $E_1$ exists if and only if $\mathcal{R}_1>1$).

At the equilibrium $E_2$, the values of $I_1,C_1,R_1$ and $Y_1$ are zero, and that of $V$ is the same as at the DFE.

The solutions of the $E_2$ system involve all  complicated square roots. In such a case,
it is more convenient to replace the ``Solve"  command by our RUR algorithm, which requires the user to input a variable to reduce $2$. The normal choice is $i_2$
(which transitions to positive at the bifurcation value), but here we will use $s$, to check the results of \cite{Bulh},  who find, using as reduction scalar
   $ x={\beta_2 j_2 } ,$
 that
\begin{eqnarray}\label{eqn:E2}
    &&s_2 =\frac{(1-\xi)\mu}{x+\mu},  \quad i_2 =\frac{(1-\xi)x\mu}{(x+\mu)(\gamma_2+\mu)}, \quad c_2 =\frac{(1-\xi)x\gamma_2\mu}{(x+\mu)(\gamma_2+\mu)(\theta_1+\mu)}, \nonumber\\ &&r_2 =\frac{(1-\xi)x\gamma_2\theta_1}{(x+\mu)(\gamma_2+\mu)(\theta_1+\mu)}, \quad z_2 =\frac{v\theta_{v}\mu}{(\theta_{v}+\mu)(\alpha_{v}x+\mu)},\\
    && y_2 =\frac{v\alpha_{v}x\theta_{v}\mu}{(\alpha_{v}x+\mu)(\theta_{v}+\mu)(\gamma_2+\mu)}\nonumber , \end{eqnarray}
and that $x$ is solution of the quadratic equation
\begin{equation}\label{eq:polinomio}
    ax^2+bx+c=0, \quad \bc a&=\alpha_{v}\nonumber\\
    b&=\mu\alpha_{v}\left[1-\frac{\beta_2(1-\xi)}{\gamma_2+\mu}\right]+\mu\left[1-\frac{\beta_2\alpha_{v}\theta_{v}v}{(\gamma_2+\mu)(\theta_{v}+\mu)}\right]\nonumber\\
    c&=\mu^2\left(1-\mathcal{R}_2\right).\ec
\end{equation}
The equilibrium $E_2$  exists  iff $\mathcal{R}_2>1$, where
\begin{equation}\label{eq:R2}
    \mathcal{R}_2=
    \frac{\beta_2}{\gamma_2+\mu}\left[1-\xi
    + \xi\frac{\alpha_{v}\theta_{v}}{\theta_{v}+\mu}\right]=
    \frac{\beta_2}{\gamma_2+\mu}\left[s_0
    + \alpha_{v}z_0 \right]=R_2(E_0)
\end{equation}
{If $\mathcal{R}_2\leq 1$, the fractions in the expression of $b$ must be smaller than one or equal to one, and it is not possible for both to be one. Therefore, $b>0$. We also have $c\geq 0$. Since that $a>0$, the equation \eqref{eq:polinomio} does not have roots with positive real parts. This implies that there is no endemic equilibrium like $E_2$. Thus, in this case, $c<0$. Since the coefficient $a$ is positive, the equation \eqref{eq:polinomio} has two real roots and only one of them is positive. In resume, if $\mathcal{R}_2>1$, there is a unique endemic equilibrium where there are infections only by the strain $2$.}

\sssec{The recipe \NGM\ $R_0$ and the  Jacobian factorization  one coincide}\la{s:RNB}

This section shows that the polynomials $R_1(X), R_2(X)$ in this example may also be obtained via the  \NGM\ approach,  as eigenvalues of the $K$ matrix,  by a judicious choice of infectious classes.

 One may choose as infectious subset the nine compartments that are $0$ in the limit, but a luckier choice here is the smaller subset $\mI=\{I_1,I_2,Y_2,Y_1\}$, which has as eigenvalues precisely the expressions $R_1(X),R_2(X)$ in \eqr{Jmax}.

The  decomposition matrices are $$\Scale[0.95]{V= \bep  \gamma_1+\mu& 0&0&0 \\
 0 & \gamma_2+\mu &0&0\\
 0&0& \gamma_1+\mu&0\\
  0&0&0& \gamma_2+\mu \eep, F=\left(
\begin{array}{cccc}
 \beta _1 s & 0 & 0 & \beta _1 s \\
 0 & \beta _2 s & \beta _2 s & 0 \\
 0 & \beta _2 z  \al _v & \beta _2 \z  \al _v & 0 \\
 0 & 0 & 0 & 0 \\
\end{array}
\right)=s B_0+ z  B_v,}$$ where
 $B_0,B_v$ are:
  \bea  B_0=\left(\begin{array}{cccc}
 \beta _1  & 0 & 0 & \beta _1  \\
 0 & \beta _2  & \beta _2  & 0 \\
 0 & 0 & 0 & 0 \\
 0 & 0 & 0 & 0 \\
\end{array}\right),\; B_v=\left(\begin{array}{cccc}
 0 & 0 & 0 & 0 \\
 0 & 0 & 0 & 0 \\
 0 & \beta _2  \al _v & \beta _2  \al _v & 0 \\
 0 & 0 & 0 & 0 \\
\end{array}\right). \eea

  The explicit non-zero eigenvalues of the  \NGM\  $(\s B_0+\\z  B_v) V^{-1}$ are  \be{R0M}\pr{ \frac{\beta _1 \s}{\gamma _1+\mu },\frac{\beta _2 \left(\\z  \al _v+\s\right)}{\gamma _2+\mu } },\ee
  confirming the result of  the Jacobian method.

  Let us note finally that \eqr{Jmax}, and the result of this section 
  imply the relation
\be{Frel} R_0=\max \pp{\mR_1, \mR_2},\ee
where $\mR_i, i=1,2$ denote the bifurcation parameters at which the boundary points
$E_i$ start to exist.

 \beR \la{r:mys}
 {Interestingly,  $R_0=\max [\mR_1,\mR_2]$  is the max of two quantities which satisfy that ${\mR_i} >1, i=1,2$ are precisely the domains where endemic points $E_i$ containing exactly one of the strains appear  -- see \eqr{Frel}.}  This formula, natural in cases where the \NGM\ has block structure, seems to be a general feature
 of multi-strain models, even when the block structure is not apparent.

 In the case of this section, there seems to be more specific structure: the Jacobian factorization approach allows introducing two ``\Dt" (see Definition \ref {d:Dt}) factors  $R_i(X), i=1,2$ of the characteristic polynomial, which are such that \BEN \im The existence conditions for $E_i$ may be expressed as $\mR_i:=R_i(DFE)>1$ -- see  \eqr{eq:R1}, \eqr{eq:R2}, \eqr{mys}.
 \im The invasion reproduction
 numbers may be obtained  simply by substituting the coordinates of the  dominance boundary equilibria
into the corresponding factor. More precisely, the invasion number of the fixed point $E_i$ for  strain $i$ is given by $R_{ji}=R_j(E_i)$.
\EEN
\eeR

{\bf Open question 2:} Does the relation $R_0=\max_1^K \mR_k $ hold  for all Descartes multi strain models of order $K$? (recall  Definition \ref{d:sms} and Lemma \ref{l:De}).


\subsubsection{The invasion reproduction number of $E_i$ is given by $R_j(E_i)$}
The invasion reproduction numbers (see for example  \cite{invasion}) may, just as the basic reproduction number, be calculated using the next generation matrix.

Our script yield quickly that
\be{invP}{\mR_{ji}} =R_j(E_i), i=1,2, j\neq i.\ee

{\bf Open question 3}: 
Do the  formulas connecting \eqr{Frel} and \eqr{invP} to the Jacobian factorization  \be{mys} \bc \mR_i =R_i(E_0), R_0=\max\pp{ \mR_1, \mR_2},\\\mR_{ji} =R_j(E_i), \text{ where $R_i$ denote polynomials obtained via}\\ \text{ the Jacobian factorization approach},\ec
 \ee
 hold, for some general class of epidemic models?

 C) For ``two strain epidemic models",  what conditions  must be satisfied to ensure the inequalities $\mR_{ji} <\mR_j, i=1,2, j\neq i$?

To resolve these questions, it might be useful to study the three and four strains generalizations of this problem, and to investigate ``non-simple" multi-strain models (in which the characteristic polynomial contains non \Dt\ polynomials).


\section{Vector-host models}
\subsection{The Jacobian $R_0$ is the square  of the  recipe NGM $R_0$ for  the dengue vector-host model without demography of \cite{brouwer2022spectral}}\la{s:Bro22}
\cite[(28)]{brouwer2022spectral} considers  a ``no demography/conservation" model with 6 compartments, three of which represent hosts, while the rest represent the vector.  Note that such models with no demography do not have a finite set of fixed points. The DFE is not unique, it coincides with the initial conditions. 
However, our algorithm works just fine. The model, after removing two ``R" classes which do not affect the rest, is:

\be{Brouw}
\bc
S_1'= -\frac{\beta_{21} I_2 S_1}{N_1}\\
S_2'= -\frac{\beta_{12} I_1 S_2}{N_2}\\
\bep I_1'\\ I_2'\eep  =\bep -\gamma_1 &\frac{\beta_{21}  S_1}{N_1}\\
 \frac{\beta_{12}  S_2}{N_2}&-\gamma_2 \eep \bep I_1\\ I_2\eep
\ec
\ee

The call ``${\text{inf}}=\{1,2\}; NGM[Brouwer22,\text{inf}]$" of our script yields that
the decomposition  matrices are
$$ F= \left(
\begin{array}{cc}
 0 & \frac{\beta_{21} S_1}{N_1} \\
 \frac{\beta_{12} S_2}{N_2} & 0 \\
\end{array}
\right), \; V= \left(
\begin{array}{cc}
 \gamma_1 & 0 \\
 0 & \gamma_2 \\
\end{array}
\right),$$
$$K=\left(
\begin{array}{cc}
 0 & \frac{\beta_{21} S_1}{\gamma_2 N_1} \\
 \frac{\beta_{12} S_2}{\gamma_1 N_2} & 0 \\
\end{array}
\right),$$
and \be{RB} R_{F}= \sqrt{ \frac{S_1 S_2 \beta_{12} \beta_{21} }{N_1 N_2 \gamma_1 \gamma_2 }}.\ee

After using {the fact that the DFE is determined by the initial conditions $S_1=N_1$, and $S_2=N_2$
we obtain  the basic reproduction number
\be{RBs} R_F=\sqrt{\frac{\beta_{12} \beta_{21}}{\gamma_1 \gamma_2}} \ee }
of \cite[40]{brouwer2022spectral}.

Here the \ch\ is of \Dt\ and the Jacobian method, and the RUR method, yield both the  square of the (modified) formula \eqr{RB} $ R_J={\frac{\beta_{12} \beta_{21}}{\gamma_1 \gamma_2}}$.

\beR  Note that \cite[35]{brouwer2022spectral}  offers yet another admissible decomposition,  based on a different  biological  interpretation, with $R_F=R_J$, and raises the question of  which of the answers is more relevant for a given epidemics.
Deciding  this from the ODE model only seems impossible.
 \eeR

\subsection{The two groups model in \cite[(5.8)]{Mart} does not obey a square relation }\la{s:Mart15}

The two groups model in  \cite[(5.8)]{Mart}  defined by 
\bea
\bc  S_1'= -\frac{\beta_{11} I_1 S_1}{N_1}-\frac{\beta_{21} I_2 S_1}{N_1}+\l_1-\mu_1 S_1\\
S_2'= -\frac{\beta_{12} I_1 S_2}{N_2}-\frac{\beta_{22} I_2 S_2}{N_2}+\l_2-\mu_2 S_2\\
\bep I_1'\\ I_2'\eep  =\bep -\gamma_1-\mu_1 -\de_1+\frac{\beta_{11}  S_1}{N_1}&\frac{\beta_{21}  S_1}{N_1}\\
 \frac{\beta_{12}  S_2}{N_2}&\frac{\beta_{22}  S_2}{N_2}-\gamma_2 -\mu_2-\de_2\eep \bep I_1\\ I_2\eep
\ec
\eea
is not anymore a vector-host model, due to the addition of the "intra-group contact infection rates" $\beta _{11},\beta _{22}$.

The  DFE is $\left\{0,0,\frac{\lambda _1}{\mu _1},\frac{\lambda _2}{\mu _2}\right\}$,
and the $R_N$ is quite complicated:
{ \bea
&&\Scale[1]{\frac{\sqrt{\left(\beta _{22} N_1 S_2 \left(\gamma _1+\delta _1+\mu _1\right)+\beta _{11} N_2 S_1 \left(\gamma _2+\delta _2+\mu _2\right)\right){}^2+4 \left(\beta _{12} \beta _{21}-\beta _{11} \beta _{22}\right) N_1 N_2 S_1 S_2 \left(\gamma _1+\delta _1+\mu _1\right) \left(\gamma _2+\delta _2+\mu _2\right)}}{2 N_1 N_2 \left(\gamma _1+\delta _1+\mu _1\right) \left(\gamma _2+\delta _2+\mu _2\right)}}\\
&&
\Scale[1]{+\frac{\beta _{22} \gamma _1 N_1 S_2+\beta _{11} \gamma _2 N_2 S_1+\beta _{22} \delta _1 N_1 S_2+\beta _{11} \delta _2 N_2 S_1+\beta _{22} \mu _1 N_1 S_2+\beta _{11} \mu _2 N_2 S_1}{2 N_1 N_2 \left(\gamma _1+\delta _1+\mu _1\right) \left(\gamma _2+\delta _2+\mu _2\right)}}.\eea}

The Jacobian factorization method yields a different answer, for a \ch\ which is not of \Dt, precisely because of the addition of $\beta _{11},\beta _{22}$.

$$R_J=\frac{\beta _{22} N_1 S_2 \left(\gamma _1+\delta _1+\mu _1\right)+\beta _{11} \gamma _2 N_2 S_1+\beta _{11} \delta _2 N_2 S_1+\beta _{11} \mu _2 N_2 S_1+\beta _{12} \beta _{21} S_1 S_2}{N_1 N_2 \left(\gamma _1+\delta _1+\mu _1\right) \left(\gamma _2+\delta _2+\mu _2\right)+\beta _{11} \beta _{22} S_1 S_2}.$$


\section{Multi-strain vector-host models}
\subsection{A  two-strain vector-host    model of Feng and Velasco-Hern\'andez \cite{feng97}, where the square relation  holds for the \brn}\la{s:Feng}

\cite{feng97} consider a human population settled in a region where a mosquito population of the genus Aedes is present and carrier of
two strains of the dengue virus. Let $V_i, I_i,  Y_i,  i=1,2$ denote the infected mosquitoes, individuals infected by one strain,    and   individuals having suffered a secondary infection, let $N=S+R + \sum_{i=1}^2 I_i+Y_i$ denote the total human population,  and let $B_1=\frac{\beta_1  V_1(t)}{c+w_h N}, B_2=\frac{\beta_2  V_2(t)}{c+w_h N}$ denote the rates of infections in human hosts produced by
 the two strains.
The model is defined as follows:
\bea
\bc
S'(t)=h-S(t)\pr{B_1+B_2}-\mu S(t),\\
I_1'(t)=B_1 S(t) -\sigma_2B_2 I_1(t) -\mu I_1(t),\\
I_2'(t)=B_2 S(t) -\sigma_1 B_1 I_2(t) -\mu I_2(t),\\
Y_1'(t)= \sigma_1 B_1 I_2(t)-(e_1+\mu +r) Y_1(t),\\
Y_2'(t)= \sigma_2 B_2 I_1(t)-(e_2+\mu +r) Y_2(t),\\
R'(t)=r(Y_1(t)+Y_2(t))-\mu R(t),\\ \\
V_1'(t)=\alpha_1 \frac{I_1(t)+Y_1(t)}{c+w_v N} M(t)-\delta V_1(t),\\
V_2'(t)=\alpha_2 \frac{I_2(t)+Y_2(t)}{c+w_v N} M(t)-\delta V_2(t)\\
M'(t)=q-M(t)\pr{\alpha_1 \frac{I_1(t)+Y_1(t)}{c+w_v N}+\alpha_2 \frac{I_2(t)+Y_2(t)}{c+w_v N}}-\delta M(t).
\ec
\eea

The  DFE is given by $E_0=\pr{h/\mu,0,0,0,0,0,0,0,q/\delta}$.
For the infectious set  $I_1,I_2,Y_1,Y_2,V_1,V_2$,  the  $F$ and $V$ matrices used in the next generation approach are given by
\bea
&&
F=\left(
\begin{array}{cccccc}
 0 & 0 & 0 & 0 & \beta_1 s_{dfe} & 0 \\
 0 & 0 & 0 & 0 & 0 & \beta_2 s_{dfe} \\
 0 & 0 & 0 & 0 & 0 & 0 \\
 0 & 0 & 0 & 0 & 0 & 0 \\
 \alpha_1 M_{dfe} & 0 & \alpha_1 M_{dfe} & 0 & 0 & 0 \\
 0 & \alpha_2 M_{dfe} & 0 & \alpha_2 M_{dfe} & 0 & 0 \\
\end{array}
\right),\\ \\ && V= \left(
\begin{array}{cccccc}
 \mu & 0 & 0 & 0 & 0 & 0 \\
 0 & \mu & 0 & 0 & 0 & 0 \\
 0 & 0 & e_1+r+\mu  & 0 & 0 & 0 \\
 0 & 0 & 0 & e_2+r+\mu  & 0 & 0 \\
 0 & 0 & 0 & 0 & \delta  & 0 \\
 0 & 0 & 0 & 0 & 0 & \delta  \\
\end{array}
\right)
\eea
with $M_{dfe}=q/\delta$.
Then,  $$F V^{-1}= \left(
\begin{array}{cccccc}
 0 & 0 & 0 & 0 & \frac{\beta_1 s_{dfe}}{{\delta}} & 0 \\
 0 & 0 & 0 & 0 & 0 & \frac{\beta_2 s_{dfe}}{\delta} \\
 0 & 0 & 0 & 0 & 0 & 0 \\
 0 & 0 & 0 & 0 & 0 & 0 \\
 \frac{\alpha_1 M_{dfe}}{{\mu}} & 0 & \frac{\alpha_1 M_{dfe}}{{e_1+\mu +\xi}} & 0 & 0 & 0 \\
 0 & \frac{\alpha_2 M_{dfe}}{\mu} & 0 & \frac{\alpha_2 M_{dfe}}{{e_2+\mu +\xi}} & 0 & 0 \\
\end{array}
\right)$$

 We obtain a basic reproduction number which is a max
\be{RFeng} \mR=max \pr{\sqrt{\mR_1},\sqrt{\mR_2}}, \mR_i:= {s_0 m_0}\frac{ {{\alpha_i \beta_i}}  }{{\delta \mu}},\ee
just like \eqr{Frel},
but contains also the extra square roots typical of vector-host models.

Furthermore, it may be checked that this is precisely the square root of the answer given by the Jacobian factorization method, which decomposes  the characteristic polynomial of the Jacobian   as the product of five linear factors with negative roots, and two quadratic \Dt\ polynomials.

There also two boundary (dominance)
equilibria, where only one strain  survives. The non-zero coordinates at the first one, $E_1$, are given by
$$\alpha_1  i_1= \delta \frac{\mR_1-1}{m_0  \beta_1/( \mu )+1  }, \beta_1 v_1= \mu \frac{\mR_1-1}{s_0 \alpha_1 (  \delta)+1  },s= \mu \frac{\alpha_1   s_0+\delta   }{\alpha_1 \beta_1 m_0+\alpha_1   \mu },$$
with similar formulas holding for the other boundary point $E_2$, by symmetry. Thus, these points
become positive precisely when  the corresponding factor  of the DFE
becomes bigger than $1$, causing instability.

Since we had trouble with computing the invasion reproduction numbers, we switched to
the ``simplified model" of \cite{feng97}, in which $M$ is eliminated by noting that  the equation for the total vector population  $T=M+V_1+V_2$ is $T'=q-\delta T$, and that, assuming  $T_0=\lim_{t\to \I} T(t)= q/\delta$,   $M$ can be removed from the system by substituting
\be{Ms} M=q/\delta -V_1-V_2.\ee

As a first consequence of using \eqr{Ms}, the $R_N$ becomes equal to $R_J$.

However, the recipe $R_0$ at $E_1$ for the natural choice of ``inf" is very complicated, and \cite{feng97} provide here a laborious local stability analysis, with complicated result, via the third order Routh-Hurwitz conditions.

We note finally that  the \ch\ for $jac(E_1)$ has two factors of degree $3$, one of which is  \Dt and one which is not. The \Dt\ factor yields a polynomial $R_1(X)$. Putting this together with its symmetric $R_2(X)$ allows finally defining
$$R_J(X)=max_j [R_1(X),R_2(X)]=max_j[\frac{\alpha_j \beta_j q s/\delta}{(\beta_j   v_j +\mu) \left( \alpha_j i_j+\delta \right)+ \beta_j   v_j \alpha_j s }].$$


\subsubsection{Invasion numbers of \cite{feng97}}
The two-strain vector-host  model in \cite{feng97} admits two boundary equilibria beside the DFE in which

 $S_1^*,S_2^*,I_1^*,I_2^*, V_1^*, V_2^*$ are the invasion infection classes.
In this case, we consider the subset $\mbox{in}_1=\pr{I_2,Y_1,Y_2,V_2}$ corresponding to the invasion infection class of $E_1$, then
\bea
&& F=\left(
\begin{array}{cccc}
 0 & 0 & 0 & b_2 S \\
 b_1 \sigma _1 v_1 & 0 & 0 & 0 \\
 0 & 0 & 0 & b_2 i_1 \sigma _2 \\
 a_2 \left(\frac{q}{\delta }-v_1\right) & 0 & a_2 \left(\frac{q}{\delta }-V_1\right) & 0 \\
\end{array}
\right),\\
&& \Scale[0.85]{V= \left(
\begin{array}{cccc}
 b_1 \sigma _1 V_1+\mu  & 0 & 0 & 0 \\
 0 & e_1+\mu +\xi  & 0 & 0 \\
 0 & 0 & e_2+\mu +\xi  & 0 \\
 a_2 \left(\frac{q}{\delta }-V_1\right)-a_2 \left(\frac{q}{\delta }-V_1-V_2\right) & 0 & a_2 \left(\frac{q}{\delta }-V_1\right)-a_2 \left(\frac{q}{\delta }-V_1-V_2\right) & a_2 \left(I_2+Y_2\right)+\delta  \\
\end{array}
\right)}, \\
&& K=\left(
\begin{array}{cccc}
 0 & 0 & 0 & \frac{b_2 S}{\delta } \\
 \frac{b_1 \sigma _1 V_1}{b_1 \sigma _1 V_1+\mu } & 0 & 0 & 0 \\
 0 & 0 & 0 & \frac{b_2 I_1 \sigma _2}{\delta } \\
 \frac{a_2 \left(\frac{q}{\delta }-V_1\right)}{b_1 \sigma _1 V_1+\mu } & 0 & \frac{a_2 \left(\frac{q}{\delta }-V_1\right)}{e_2+\mu +\xi } & 0 \\
\end{array}
\right)
\eea
 then the IRN of strain 1 at $E_1$ is
 $$ R_1=\frac{\sqrt{a_2} \sqrt{b_2} \sqrt{q} \sqrt{S \left(e_2+\mu +\xi \right)}}{\delta  \sqrt{\mu } \sqrt{e_2+\mu +\xi }}.$$\\

 Similarly, we chose the other subset  $\mbox{in}_1=\pr{I_1,Y_1,Y_2,V_1}$ corresponding to the invasion infection class at $E_2$, we obtain
 \bea
&& F= \left(
\begin{array}{cccc}
 0 & 0 & 0 & b_1 S \\
 0 & 0 & 0 & b_1 I_2 \sigma _1 \\
 b_2 \sigma _2 V_2 & 0 & 0 & 0 \\
 a_1 \left(\frac{q}{\delta }-V_2\right) & a_1 \left(\frac{q}{\delta }-V_2\right) & 0 & 0 \\
\end{array}
\right),\\
&& \Scale[0.85]{V= \left(
\begin{array}{cccc}
 b_2 \sigma _2 V_2+\mu  & 0 & 0 & 0 \\
 0 & e_1+\mu +\xi  & 0 & 0 \\
 0 & 0 & e_2+\mu +\xi  & 0 \\
 a_1 \left(\frac{q}{\delta }-V_2\right)-a_1 \left(\frac{q}{\delta }-V_1-V_2\right) & a_1 \left(\frac{q}{\delta }-V_2\right)-a_1 \left(\frac{q}{\delta }-V_1-V_2\right) & 0 & a_1 \left(I_1+Y_1\right)+\delta  \\
\end{array}
\right)},\\
&& K=\left(
\begin{array}{cccc}
 0 & 0 & 0 & \frac{b_1 S}{\delta } \\
 0 & 0 & 0 & \frac{b_1 I_2 \sigma _1}{\delta } \\
 \frac{b_2 \sigma _2 V_2}{b_2 \sigma _2 V_2+\mu } & 0 & 0 & 0 \\
 \frac{a_1 \left(\frac{q}{\delta }-V_2\right)}{b_2 \sigma _2 V_2+\mu } & \frac{a_1 \left(\frac{q}{\delta }-V_2\right)}{e_1+\mu +\xi } & 0 & 0 \\
\end{array}
\right)
 \eea
 then the maximum eigenvalue of K yields the IRN at $E_2$ which is
 $$ R_2=\frac{\sqrt{a_1} \sqrt{b_1} \sqrt{q-\delta  v_2} \sqrt{I_2 \sigma _1 \left(b_2 \sigma _2 V_2+\mu \right)+e_1 S+S (\mu +\xi )}}{\delta  \sqrt{e_1+\mu +\xi } \sqrt{b_2 \sigma _2 V_2+\mu }}.$$

\ssec{The Dengue- Zika model with  coinfection  and ADE \cite{KriOla}}\la{s:KriOla}
The model studied in this paper continues previous papers like
Isea and Lonngren 2016 \cite{Isea16}, and Okuneye et al. 2017 \cite{Okuneye17},  most notably by taking into account the possibility of coinfection and of direct transmission of Zika via sex (which entails two forces of infection for Zika transmissions in their flow-chart, and hence an asymmetry in the results).

 Introduce  the following forces of infection:
\be{Fo} \bc F_{vd}=\beta _{hd} T_{vd}, T_{vd}=I_{vd}+I_{ {vc}} \nu _d, & \text{ dengue  vector force}\\
F_{vz}=\beta _{hz} T_{vz}, T_{vz}=I_{vz}+I_{ {vc}} \nu _z, & \text{ zika vector force}
\\ F_{hz}=\beta _{vz} T_{hz}, T_{hz}=I_z+I_c+J_z k_z, & \text{ zika human force}
\\ F_{hd}=\beta _{vd} T_{hd}, T_{hd}=I_d+I_c+J_d k_d,  & \text{ dengue human force}\\F_s=\beta _s T_{hz}& \text{ zika human to human force}.\ec \ee
Note that $\nu _d, \nu _z$  and $k_d,k_z$ are respectively  the parameters of altered infectivity for co-infected vectors  and of ADE, and  note that even when  $\nu _d=\nu _z=1$, the co-infection model is  more accurate than  previous works like \cite{feng97}, since it takes into account the existence of  doubly infected vectors $I_{ {vc}}$ which influence both chains of infection.

We will consider   the model :
\be{ZD}  \bc S_h'=(N_h-S_h)  \mu -S_h \left(F_{vd} +F_{vz} +
F_{s} \right),\\
I_d'= S_h F_{vd}-\rho I_{d} \left(F_{vz} +F_{s}\right) -I_d \left(\g_d+\mu\right),\\
I_z'= S_h (F_{vz}+F_s)-\rho I_{z} F_{vd}-I_z \left( \g_d+\mu\right),\\
I_c'=\rho \pp{I_{d} \left(F_{vz} +F_{s}\right) + I_{z} F_{vd} -I_c \left( \g_d+\g_c\right)}- \mu I_c,\\
R_d'=I_d \gamma _d-R_d  \left(F_{vz} +F_{s} +\mu\right),\\
R_z'=I_z \gamma _d-R_z  \left(F_{vd} +\mu\right),\\
J_d'=\rho \g_z I_c+ R_z  \left(F_{vd} -\g_d-\mu\right),\\
J_z'=\rho \gamma _d I_c +R_d  \left(F_{vz} + F_{s} -\g_z-\mu\right),\\
R'=J_d \gamma _d+J_z \gamma _z-\mu  R,\\
S_v'=(N_v -S_v) \mu _v-S_v ( F _{hd}+ T_{hz}),\\
I_{vd}'
= {F _{hd} S_v -\rho F _{hz} I_{vd}  -I_{vd}  \mu _v},\\
I_{vz}'= {F _{hz} S_v -\rho F _{hd} I_{vz}-I_{vz}  \mu _v},\\
I_{vc}'=\rho \pr{F _{hz} I_{vd}  + F _{hd} I_{vz}}-I_{ {vc}}  \mu _v,\ec
\ee
which  generalizes a bit \cite{KriOla} by introducing the parameter $\rho$, whose purpose is to allow simplifying the model to remove the $I_c, I_{vc}$ classes, by setting $\rho=0$.

Note that humans are born fully susceptible to dengue and Zika at a rate of $            \mu {N_h}$, where $\mu$ is the
natural birth/death rate for humans and $N_h$ is the total human population. Susceptible
individuals can become infected with dengue from either a dengue-infected ($I_{vd}$) or
coinfected female mosquito ($I_{vc}$). The mosquito-to-human dengue infection rate is
given by $\b_{hd}$. This rate is modified by a factor of $\nu_d$ to indicate the altered infectivity
of coinfected mosquitoes. Once infected with dengue, humans can recover or become
co-infected with Zika (by a Zika-infected ($I_{vz}$) or coinfected female mosquito ($I_{vc}$),
or by sexual transmission from a Zika-infected ($I_z$) or coinfected ($I_c$) human) and
transition into the Rd or Ic class, respectively. In a similar manner, fully susceptible
humans become infected with Zika from a mosquito in the $I_{vz}$ or $I_{vc}$ compartment.

The DFE has only non-zero components $S_v=N_v,S_h=N_h$. Choosing
as infectious set all the \com s except $S_v,S_h$
yields
\be{RKri}R_0=\max[\sqrt{\frac{\beta _{ {hd}} N_v \beta _{ {vd}}}
{N_h \mu _v(\gamma _d +\mu )}}, \frac{\beta _s+ \sqrt{\beta _s^2+\frac{4 N_v\beta _{ {hz}}  \beta _{ {vz}} \left(\mu +\gamma _z\right)}{N_h \mu _v}}}{2 \left(\mu +\gamma _z\right)}]:=\max[\mR_d, \mR_z],\ee
confirming \cite[sec. 4]{KriOla}, and also the multi-strain structure we already
met in \eqr{Frel}, \eqr{RFeng}. Furthermore, one may show that $\mR_d>1, \mR_z>1$ are  necessary and sufficient conditions for the existence of the dengue only and Zika only fixed points -- see subsequent sections.

We end this section by reporting  on the Jacobian factorizations at $E_0$, when choosing
as infectious set $$\mI=\left\{I_d,I_z,I_c,J_d,J_z,I_{ {vd}},I_{ {vz}},I_{ {vc}}\right\}.$$
Now the \ch\ has two second order factors:
\BEN \im  One of \Dt\ which yields  the polynomial
$ R_1(X)=\frac{\beta _{ {hd}} S_v \beta _{ {vd}} \left(k_d R_z+S_h\right)}{N_h^2 \mu _v \left(\gamma _d+\mu \right)},$ which generalizes $\mR_d$, in the sense  that $ R_1(E_0)=\mR_d^2$; this  {raises the question of whether this is related to the Zika IRN.}
\im One not of \Dt,  which   {raises the question of how to exploit non \Dt\ second order factors}.
\EEN

\sssec{The dengue only resident fixed point $E_d$}
Even though the coordinates of the dengue only resident fixed point $E_d$ are pretty simple,
obtaining them isn't.  We have an a priori choice of zeroable set
$ in_{1'}=\{I_z,R_z,J_z,I_{zv}\}$ which turns out to lead to about 2.5 hrs for "Solve" (due to the existence of 4 extra fixed points, which are non-positive for the numeric values of \cite{KriOla}. After performing the computation, it turns out that the full zeroable set is $in_1 = \{I_z,I_c,R_z,J_d,J_z,R,I_{vz},I_{vc}\}$.  The remaining set of equations:

$$\left(
\begin{array}{cc}
 -\gamma _d I_d-\mu  I_d+\frac{I_{ {vd}} S_h \beta _{ {hd}}}{N_h}&=0 \\
 \gamma _d I_d-\mu  R_d &=0\\
 \frac{I_d S_v \beta _{ {vd}}}{N_h}-I_{ {vd}} \mu _v &=0\\
 \mu  \left(N_h-S_h\right)-\frac{I_{ {vd}} S_h \beta _{ {hd}}}{N_h}&=0 \\
 -\frac{I_d S_v \beta _{ {vd}}}{N_h}+N_v \mu _v-S_v \mu _v &=0\\
\end{array}
\right)$$
may be easily solved.  Besides the DFE, it has one extra fixed point:
\bea && R_d= \frac{\gamma _d I_d}{\mu },
S_v=\frac{\mu _v   N_h N_v}
{ \mu _v N_h+  \beta _{ vd} I_d}=\frac{\mu _v \left(\gamma _d+\mu \right) \left(\mu  N_h+\beta _{ {hd}} N_v\right)}
{\beta _{ {hd}}\pp{ \mu _v \left(\gamma _d+\mu \right)+\mu   \beta _{ vd}}},\\&&S_h=\frac{N_h^2 \left(\mu _v \left(\gamma _d+\mu \right)+\mu  \beta _{ {vd}}\right)}{\beta _{ {vd}} \left(\mu  N_h+\beta _{ {hd}} N_v\right)},
I_d=\frac{\mu  N_h^2\mu _v  }{\beta _{ {vd}}  \left(\mu  N_h+\beta _{{hd}} N_v\right)}
\left(\fr{N_v\beta _{{hd}}  \beta _{ {vd}}}{N_h \mu _v \left(\gamma _d+\mu \right)}-1\right),\\
&& I_{{dv}}= \beta _{ vd} I_d \frac
{   S_v}{\mu _v   N_h}= \frac{I_d N_v \beta _{ {vd}}}{I_d \beta _{ {vd}}+N_h \mu _v},I_{{vz}}= 0,I_{vc}= 0.
\eea

The bifurcation value for $E_d$ is thus
$$ \fr{N_v\beta _{{hd}}  \beta _{ {vd}}}{N_h \mu _v \left(\gamma _d+\mu \right)}:=\mR_d^2,$$
confirming \cite[Lem. 1]{KriOla}.

The Jacobian factorizations  when choosing
as infectious set the  complement of $I_d,R_d,I_{ {vd}},S_v,S_h$
has \ch\ with one non-\Dt,  third order factor.

\sssec{The Zika only resident fixed point $E_z$}
Using  the full zeroable set given in \cite{KriOla} $in_2 = \{I_d,I_c,R_d,J_d,J_z,R,I_{dv},I_{vc}\}$, yields   the  set of equations:
$$\left(
\begin{array}{cc}
 S_h \left(\frac{\beta _{ {hz}} I_{ {vz}}}{N_h}+\frac{I_z \beta _s}{N_h}\right)-I_z \gamma _z-\mu  I_z &=0\\
 I_z \gamma _z-\mu  R_z &=0\\
 \frac{I_z S_v \beta _{ {vz}}}{N_h}-I_{ {vz}} \mu _v &=0\\
 \mu  \left(N_h-S_h\right)-S_h \left(\frac{\beta _{ {hz}} I_{ {vz}}}{N_h}+\frac{I_z \beta _s}{N_h}\right) &=0\\
 -\frac{I_z S_v \beta _{ {vz}}}{N_h}+N_v \mu _v-S_v \mu _v &=0\\
\end{array}
\right)$$
The Zika only resident fixed point $E_z$ satisfies
\bea && R_z= \frac{\gamma _z I_z}{\mu },
S_v=\frac{\mu _v   N_h N_v}
{ \mu _v N_h+  \beta _{ vz} I_z},
I_{{dv}}= \beta _{ vz} I_z \frac
{   S_v}{\mu _v   N_h}= \frac{I_z  \beta _{ {vz}} N_v}{ \mu _v N_h+  \beta _{ vz} I_z},I_{{vz}}= 0,I_{vc}= 0, \\&&S_h=\frac{\mu  N_h^2 \left(N_h \mu _v+I_z \beta _{{vz}}\right)}{I_z \beta _{{vz}} \left(\mu  N_h+\beta _{{hz}} N_v+I_z \beta _s\right)+N_h \mu _v \left(\mu  N_h+I_z \beta _s\right)}=\frac{N_h^2 \left(\mu _v \left(\gamma _d+\mu \right)+\mu  \beta _{ {vd}}\right)}{\beta _{ {vd}} \left(\mu  N_h+\beta _{ {hd}} N_v\right)}
, \eea
where
$I_z$ is a positive root of the  quadratic equation $a I_z^2 + b I_z +c=0$, with coefficients:

$$\bc c=\mu  N_h \left(N_h \mu _v \left(\mu -\beta _s+\gamma _z\right)-\beta _{ {hz}} N_v \beta _{ {vz}}\right),\\
b=N_h \beta _s \mu _v \left(\mu +\gamma _z\right)+\mu  N_h \beta _{ {vz}} \left(\mu -\beta _s+\gamma _z\right)+\beta _{ {hz}} N_v \beta _{ {vz}} \left(\mu +\gamma _z\right),\\
a=\beta _s \beta _{ {vz}} \left(\mu +\gamma _z\right) \ec$$

Assume first that $\b_s$ is small enough so that $b>0$; then, this equation has a unique positive root iff $c<0$, which may be written also as
\be{my}  \frac{N_h \beta _s \mu _v+\beta _{ {hz}} N_v \beta _{ {vz}}}{N_h \mu _v(\mu + \gamma _z)} >1.\ee

It is shown in  \cite[Thm 1]{KriOla} that this is equivalent  to $R_z >1$ (both conditions determine the correct stability domain, and both reduce when $\b_s=0$ to the same answer $\frac{\beta _{ {hz}} N_v \beta _{ {vz}}}{ N_h \mu _v(\mu + \gamma _z)}$).

The model of \cite{KriOla} contains several interesting particular cases, to which we turn next. 
\sssec{The dengue invasion reproduction number (IRN) and two possible $(F,V)$ decompositions}\label{s:KriOlaIRN}
The dengue fixed point has non-zero values $S_h,S_v,I_d,R_d, I_{dv}$. Computing the IRN's requires specifying the "invasion infection classes".
\cite{KriOla} work with a subset of $$in_{2'} = \{I_d,I_c,R_d,J_d,J_z,I_{dv},I_{cv},R_c\},$$
given by $in_2 = \{I_d,I_c,J_d,I_{dv},I_{cv}\}$.

The  resulting recipe $V$ matrix is diagonal, and the recipe $F$ matrix, after denoting proportions by minuscule letters,  is:
\be{FKrrec}F=\left(
\begin{array}{ccccc}
 0 & 0 & 0 &  {s_h \beta _{ {hd}}}  &  {\nu _d s_h \beta _{ {hd}}}  \\
  {\rho  \left(\beta _{ {hz}} i_{ {zv}}+i_z \beta _s\right)}  & 0 & 0 & {\rho  \beta _{ {hd}} i_z}  &  {\rho  \nu _d \beta _{ {hd}} i_z}  \\
 0 & 0 & 0 &  {\beta _{ {hd}}r_z}  &  {\nu _d \beta _{ {hd}}r_z}  \\
  {s_v \beta _{ {vd}}}  &  {s_v \beta _{ {vd}}}  &  {k_d s_v \beta _{ {vd}}}  & 0 & 0 \\
  {i_{ {zv}} \beta _{ {vd}}}  &  {i_{ {zv}} \beta _{ {vd}}}  &  {k_d i_{ {zv}} \beta _{ {vd}}}  &  {i_z \beta _{ {vz}}}  & 0 \\
\end{array}
\right)\ee
and the spectral radius  of the resulting recipe $K$ matrix satisfies a polynomial equation of degree $4$.

 Now \cite[Sec 5.1]{KriOla} move two of the $F$ terms in the $V$ matrix,
 yielding
 \be{FKr}F=\left(
\begin{array}{ccccc}
 0 & 0 & 0 &  {s_h \beta _{ {hd}}}  &  {\nu _d s_h \beta _{ {hd}}}  \\
  0  & 0 & 0 & {\rho  \beta _{ {hd}} i_z}  &  {\rho  \nu _d \beta _{ {hd}} i_z}  \\
 0 & 0 & 0 &  {\beta _{ {hd}}r_z}  &  {\nu _d \beta _{ {hd}}r_z}  \\
  {s_v \beta _{ {vd}}}  &  {s_v \beta _{ {vd}}}  &  {k_d s_v \beta _{ {vd}}}  & 0 & 0 \\
  {i_{ {zv}} \beta _{ {vd}}}  &  {i_{ {zv}} \beta _{ {vd}}}  &  {k_d i_{ {zv}} \beta _{ {vd}}}  &  0  & 0 \\
\end{array}
\right),\ee
with the $-V$ matrix being:

\be{VKr} \left(
\begin{array}{ccccc}
 -\gamma _d-\mu-{\rho  \left(\beta _{ {hz}} i_{ {zv}}+i_z \beta _s\right)}  & {\rho  \left(\beta _{ {hz}} i_{ {zv}}+i_z \beta _s\right)} & 0 & 0 & 0 \\
 0 & -\rho  \left(\gamma _d+\gamma _z\right)-\mu  &  \rho  \gamma _z & 0 & 0 \\
 0 &0 & -\gamma _d-\mu  & 0 & 0 \\
 0 & 0 & 0 & -{i_z \beta _{ {vz}}}-\mu _v & {i_z \beta _{ {vz}}} \\
 0 & 0 & 0 & 0 & -\mu _v \\
\end{array}
\right)\ee
They reduce thus the rank of $K$ to $2$ and getting a simpler $R_0$. On the other hand, their decomposition is admissible only under extra conditions of the parameters which ensure the non-positivity of the row-sums of $-V$, which they omit to mention.

\beR \la{r:VFKr} The associated CTMC is the union of two disjoint generalized Erlangs, on the host and vector, respectively. These are employed in the probabilistic/epidemic interpretations in \cite{KriOla}.

The probabilistic/epidemic significance of $F$ is better understood after decomposing this matrix as a sum of matrices of rank $1$ as follows:
\be{FKrdec}F=\bep
  {\beta _{ {hd}}} s_h   \\   {\rho  \beta _{ {hd}} i_z}  \\
   {\beta _{ {hd}}k_d r_z}    \\
   0  \\
  0
\eep
\bep 0&0&0&1& {\nu _d }\eep +
\bep 0\\0\\0\\
  { \beta _{ {vd}}} s_v \\
  { \beta _{ {vd}}} i_{ {zv}} \eep \bep 1&1&1&0&0\eep. \ee

   The column vector are total infectivity rates for the resident compartments,  the row vectors are distribution vectors, and this decomposition yields immediately both the Diekmann kernel and $R_0$ -- see \cite{AABBGH,AAGH}.
   \eeR


 \sec{Conclusions}

 The possible non-uniqueness of the NGM matrix has not been sufficiently studied in the literature. Sometimes, like in the example of the last section, one simplifying choice is justified a posteriori  on the grounds of some interpretability of the results, ignoring the fact that other choices might lead to even simpler answers, and the fact that a priori there is no reason to expect simple answers.

 We answer to this classic dilemma by  showing via numerous examples that the first ``recipe NGM" to come to mind   leads quickly
 to most of the  results found in the literature. The question of whether our recipe may  always  be associated to admissible  equation decompositions remains open.

  We have also examined a variant of the Jacobian approach, a "factorization Jacobian approach", which draws the attention to certain polynomials with interesting properties \eqr{mys},
  and raises interesting questions  -- see especially Section 4.2, Open Question 3. Notably, the   relation
\eqr{Frel}   holds in all the three "multi-strain" examples we examined, and raises the additional question of how to define multi-strain models in terms of the dynamical system, to ensure that this always holds for this class.

 \sec{Appendix: the implementation  of the  Jacobian factorization approach} \la{s:Jac}
First, we use a utility which, for a given model, infectious set, and dummy variable (taken always as $u$, to avoid confusions), outputs the Jacobian at the DFE, the  trace
and determinant (for other purposes), the characteristic polynomial in $u$, the NGM matrix  and $R_F$.
\begin{verbatim}
JR0[mod_,inf_,u_,cn_:{}]:=
  Module[{dyn,X,par,cinf,cp,cX,jac,tr,det,chp,ngm,K,R0},
    dyn=mod[[1]];X=mod[[2]];par=mod[[3]];
    Print[" dyn=",dyn//FullSimplify//MatrixForm,X,par];
    cinf=Thread[X[[inf]]->0];
    cp=Thread[par>0];cX=Thread[X>0];
    cdfe=Join[DFE[mod,inf],cinf];
    jac=Grad[dyn,X]/.cinf/.cn;
    tr=Tr[jac];
    det=Det[jac];
    chp=CharacteristicPolynomial[jac,u];
    ngm=NGM[mod,inf];
    K=ngm[[6]];
   Print["K=",K//MatrixForm];
   R0=Assuming[Join[cp,cX],Max[Eigenvalues[K]]];
  {chp,R0,K,jac,tr,det}];
\end{verbatim}

Most of the work is done after calling this utility, by another one, JR02.This splitting of JR0 in two parts is necessary  since
the detection of the non-sign definite factors which must be analyzed  is easier to perform by eye, than to program. The JR02  script is:
\begin{verbatim}
JR02[pol_,u_]:=Module[{co,co1,cop,con,R_J},co=CoefficientList[pol,u];
  Print["the  factor ",pol," has degree ",Length[co]-1];
  co1=Expand[co[[1]]* co[[Length[co]]]];
  Print["its leading * constant coefficient product is ",co1];
  cop=Replace[co1, _. _?Negative -> 0, {1}](*level 1 here ?*);
  con=cop-co1;
  Print["R_J is"];
  R_J=con/cop//FullSimplify;
{R_J,co}
]
\end{verbatim}

For a specific ``mod", both $R_0$'s may be obtained by typing:
\begin{verbatim}
jr = JR0[mod, inf, u];
chp = jr[[1]] // Factor
Print["factor is ", pol = chp[[5]]]
pc = JR02[pol,
   u];(*the script JR02 determines R_J, using the index, 
   for example 5, determined by \eye inspection in the previous command*)
Print["R_J is ", R_J = pc[[1]] // FullSimplify]
Print["R_N is ", R_N = jr[[2]] // FullSimplify]
\end{verbatim}

 \subsection*{Proof of \cite{whittle1955outcome}'s result via Mathematica}\la{s:Whi}
\BEN
\im  The solution of the first recurrence equation in \eqr{recS} for the expected time to extinction of a linear birth and death process with arrival rate $A$ and death rate $q A$ (relevant when $R_0<1$), via Mathematica is:
\bea
\Scale[0.85]{\frac{q \left(H_K \left(1-q^j\right)+H_j \left(q^K-1\right)+\log \left(\frac{q-1}{q}\right) \left(q^K-q^j\right)\right)-\left(\left(q^j-1\right) \Phi \left(\frac{1}{q},1,K+1\right)\right)+\left(q^K-1\right) \Phi \left(\frac{1}{q},1,j+1\right)}{A (q-1) q \left(q^K-1\right)}},
\eea
where $H$ denotes the Harmonic function.

Since Mathematica cannot compute the limit when $K$ converges to infinity directly,
we break the limit into  its  three parts, and end up with the following generalization:

 Making now $j= 1$ yields \cite{whittle1955outcome}'s result which is
$$ \frac{\log (q)-\log (q-1)}{A}.$$

\im When $R_0>1$ we cannot obtain the limit  for general $j$. When $j = 1$,  similarly with the previous case,  the limit   is devided into four parts:
\bea
\Scale[0.85]{\bc
a_1=\text{Limit}\left[\frac{\frac{q \left(q^K \left(q \left(-\left(-\frac{\log (1-q)}{q}-1\right)\right)\right)\right)}{q}-q^K \left(H_K+\log (1-q)\right)}{A (q-1) \left(q^K-1\right)},K\to \infty ,\text{Assumptions}\to \{A>0,0<q<1\}\right],\\
a_2=\text{Limit}\left[\frac{q \left(\left(H_K-1\right) q^K+\log (1-q)-\frac{\log (1-q)}{q}\right)}{A (q-1) \left(q^K-1\right)},K\to \infty ,\text{Assumptions}\to \{A>0,0<q<1\}\right],\\
a_3= \text{Limit}\left[-\frac{q q^K \Phi (q,1,K+1)}{A (q-1) \left(q^K-1\right)},K\to \infty ,\text{Assumptions}\to \{A>0,0<q<1\}\right],\\
a_4=\text{Limit}\left[\frac{q \left(q^K (q \Phi (q,1,K+1))\right)}{A (q-1) \left(q^K-1\right)},K\to \infty ,\text{Assumptions}\to \{A>0,0<q<1\}\right]
\ec}
\eea

{Here Mathematica yields that $a_1 = 0$, $a_2= -\frac{\log (1-q)}{A}$, the second being precisely Whittle's result, but we were unable to confirm with Mathematica} that $a_3=a_4=0$.

\EEN 

  {\bf Acknowledgements}. We thank
 Andrew Brouwer,  Corey Shanbrom, Matija Vidmar, Dan Goreac, Andrei Halanay and James Watmough for useful exchanges.

  \bibliographystyle{plain}
\bibliography{Pare40}

\end{document}